\documentclass[%
 reprint,
superscriptaddress,
showpacs,
nofootinbib,
 amsmath,amssymb,
 aps,
prc,
floatfix,
]{revtex4-1}

\usepackage{graphicx}% Include figure files
\usepackage{dcolumn}% Align table columns on decimal point
\newcolumntype{.}{D{.}{.}{-1}}
\usepackage{bm}% bold math
\usepackage{float,array,booktabs,amssymb}
\begin{document}

\preprint{APS/123-QED}

\title{
%High precision branching ratio measurements in $^{19}$Ne $\beta^+$ decay
%The ${\cal F}t$ value of $^{19}$Ne beta decay and constraints on new physics
Precise branching ratio measurements in $^{19}$Ne beta decay and fundamental tests of the weak interaction
}% Force line breaks with \\
\author{B.\,M.~Rebeiro}
\affiliation{Department of Physics and Astronomy, University of the Western Cape, P/B X17, Bellville 7535, South Africa}
\author{S.~Triambak}
\email{striambak@uwc.ac.za}
\affiliation{Department of Physics and Astronomy, University of the Western Cape, P/B X17, Bellville 7535, South Africa}
%\affiliation{iThemba LABS, P.O. Box 722, Somerset West 7129, South Africa}
\author{P.\,Z.~Mabika}
\affiliation{Department of Physics and Astronomy, University of the Western Cape, P/B X17, Bellville 7535, South Africa}
\affiliation{Department of Physics, University of Zululand, Private Bag X1001, KwaDlangezwa 3886, South Africa}
\author{P.~Finlay} 
\affiliation{Department of Physics, University of Guelph, Guelph, Ontario N1G 2W1, Canada}
% \author{R.~Dunlop} 
% \affiliation{Department of Physics, University of Guelph, Guelph, Ontario N1G 2W1, Canada}
\author{C.\,S.~Sumithrarachchi} 
\altaffiliation[Present address: ]{National Superconducting Cyclotron Laboratory, Michigan State University, East Lansing, Michigan 48824, USA.}
\affiliation{Department of Physics, University of Guelph, Guelph, Ontario N1G 2W1, Canada}
\author{G.~Hackman} 
\affiliation{TRIUMF, 4004 Wesbrook Mall, Vancouver, British Columbia V6T 2A3, Canada}
\author{G.\,C.~Ball} 
\affiliation{TRIUMF, 4004 Wesbrook Mall, Vancouver, British Columbia V6T 2A3, Canada}
\author{P.\,E.~Garrett}
\affiliation{Department of Physics, University of Guelph, Guelph, Ontario N1G 2W1, Canada}
\author{C.\,E.~Svensson} 
\affiliation{Department of Physics, University of Guelph, Guelph, Ontario N1G 2W1, Canada}
\author{D.\,S.~Cross}
%\affiliation{TRIUMF, 4004 Wesbrook Mall, Vancouver, British Columbia V6T 2A3, Canada}
\affiliation{Department of Chemistry, Simon Fraser University, Burnaby, British Columbia V5A 1S6, Canada}
\author{R.~Dunlop} 
\affiliation{Department of Physics, University of Guelph, Guelph, Ontario N1G 2W1, Canada}
\author{A.\,B.~Garnsworthy} 
\affiliation{TRIUMF, 4004 Wesbrook Mall, Vancouver, British Columbia V6T 2A3, Canada}
\author{R.~Kshetri} 
\altaffiliation[Present address: ]{Department of Physics, The University of Burdwan, Burdwan 713104, West Bengal, India}
\affiliation{TRIUMF, 4004 Wesbrook Mall, Vancouver, British Columbia V6T 2A3, Canada}
\affiliation{Department of Chemistry, Simon Fraser University, Burnaby, British Columbia V5A 1S6, Canada}
\author{J.\,N.~Orce} 
%\affiliation{TRIUMF, 4004 Wesbrook Mall, Vancouver, British Columbia V6T 2A3, Canada}
\affiliation{Department of Physics and Astronomy, University of the Western Cape, P/B X17, Bellville 7535, South Africa}
\author{M.\,R.~Pearson} 
\affiliation{TRIUMF, 4004 Wesbrook Mall, Vancouver, British Columbia V6T 2A3, Canada}
\author{E.\,R.~Tardiff} 
\altaffiliation[Present address: ]{Department of Physics and Astronomy, Northwestern University, Evanston, IL 60208, USA}
\affiliation{TRIUMF, 4004 Wesbrook Mall, Vancouver, British Columbia V6T 2A3, Canada}
\author{H.~Al-Falou} 
\affiliation{TRIUMF, 4004 Wesbrook Mall, Vancouver, British Columbia V6T 2A3, Canada}
\author{R.\,A.\,E.~Austin} 
\affiliation{Department of Astronomy \& Physics, Saint Mary's University, Halifax, Nova Scotia B3H 3C3, Canada}
\author{R.~Churchman}
\altaffiliation[Deceased.]{}
\affiliation{TRIUMF, 4004 Wesbrook Mall, Vancouver, British Columbia V6T 2A3, Canada}
\author{M.\,K.~Djongolov} 
\affiliation{TRIUMF, 4004 Wesbrook Mall, Vancouver, British Columbia V6T 2A3, Canada}
\author{R.~D'Entremont} 
\affiliation{Department of Astronomy \& Physics, Saint Mary's University, Halifax, Nova Scotia B3H 3C3, Canada}
\author{C.~Kierans} 
\affiliation{Physics Department, Simon Fraser University, Burnaby, British Columbia V5A 1S6, Canada}
\author{L.~Milovanovic} 
\affiliation{Department of Physics \& Astronomy, University of British Columbia, Vancouver, British Columbia V6T 1Z4, Canada}
\author{S.~O'Hagan} 
\affiliation{Department of Science, University of Alberta Augustana Campus, Camrose, Alberta T4V 2R3, Canada}
\author{S.~Reeve} 
\affiliation{Department of Astronomy \& Physics, Saint Mary's University, Halifax, Nova Scotia B3H 3C3, Canada}
\author{S.\,K.\,L.~Sjue}
\altaffiliation[Present address: ]{Physics Division, Los Alamos National Laboratory, Los Alamos, NM 87545, USA}
\affiliation{TRIUMF, 4004 Wesbrook Mall, Vancouver, British Columbia V6T 2A3, Canada}
\author{S.\,J.~Williams} 
%\altaffiliation[Present address: ]{National Superconducting Cyclotron Laboratory, Michigan State University, East Lansing, Michigan 48824, USA.}
\affiliation{TRIUMF, 4004 Wesbrook Mall, Vancouver, British Columbia V6T 2A3, Canada}
 \author{S.\,S.~Ntshangase}
 \affiliation{Department of Physics, University of Zululand, Private Bag X1001, KwaDlangezwa 3886, South Africa}
\date{\today}% It is always \today, today,

\begin{abstract}
We used the 8$\pi$ $\gamma$-ray spectrometer at the TRIUMF-ISAC radiocative ion beam facility to obtain high-precision branching ratios for $^{19}$Ne $\beta^+$ decay to excited states in $^{19}$F. Together with other previous work, our measurements determine the superallowed $1/2^+ \to 1/2^+$ beta branch to the ground state in $^{19}$F to be 99.9878(7)\%, which is three times more precise than known previously. The implications of these measurements for testing a variety of  weak interaction symmetries are discussed briefly.
%% \begin{description}
%% \item[Background]       
%% \item[Purpose]
%% \item[Methods]
%% \item[Results]
%% \item[Conclusions]
%% \end{description}
\end{abstract}

\maketitle

\section{Introduction}
High precision measurements of observables in $^{19}$Ne $\beta^+$ decay offer several opportunities to rigorously test symmetries of the weak interaction.
%that form an integral part of the Standard Model.  
For example, correlation measurements from the decay have been previously used to search for second-class~\cite{calaprice:75} and right-handed weak interactions~\cite{carnoy:92,holstein:77}, as well as set stringent limts on Fierz interference terms~\cite{holstein:fierz} and time-reversal-odd currents~\cite{hallin:84,schneider:83}. Such experiments have constituted valuable probes for physics beyond the standard model (BSM). In addition, precision measurements of $^{19}$Ne $\beta^+$ decay transition probabilities provide a test of the conserved vector current (CVC) hypothesis, allowing a determination of $V_{ud}$, the up-down element of the Cabibbo-Kobayashi-Maskawa (CKM) quark mixing matrix~\cite{oscar:09} and are also important to test shell model calculations~\cite{haxton:80,brown:80} used to interpret parity mixing in $^{19}$F~\cite{ega:83,Adelberger:85}.
% 
% stringently tested the time-reversal invariant nature of the charged weak interaction. In addition, measured $^{19}$Ne $\beta^+$ decay rates allow a determination of $V_{ud}$, the up-down element of the Cabibbo-Kobayashi-Maskawa (CKM) quark mixing matrix~\cite{oscar:09} and also provide an important test of shell model calculations~\cite{haxton:80,brown:80} that are used to interpret parity mixing in $^{19}$F~\cite{ega:83}. 

\begin{figure}[t]
\includegraphics[scale=0.62]{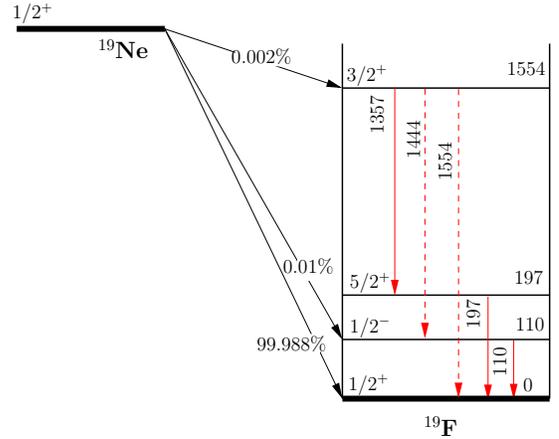}% Here is how to import EPS art
\caption{\label{fig:levels}Nominal decay scheme and $\gamma$ transitions for $^{19}$Ne $\beta^+$ decay. The decay proceeds predominantly via the superallowed $1/2^+ \to 1/2^+$ transition to the ground state in $^{19}$F. The solid red lines represent the most intense $\gamma$ transitions from excited states in $^{19}$F. Level energies are in keV.}
\end{figure}

\noindent In this paper, we report precise measurements of $^{19}$Ne $\beta^+$ decay branches to excited states in $^{19}$F, shown in Fig.~\ref{fig:levels}. We briefly discuss the implications of our results for fundamental tests of the weak interaction.  

\section{Experimental details}
\subsection{Apparatus}
\begin{figure}[t]
\includegraphics[scale=0.33]{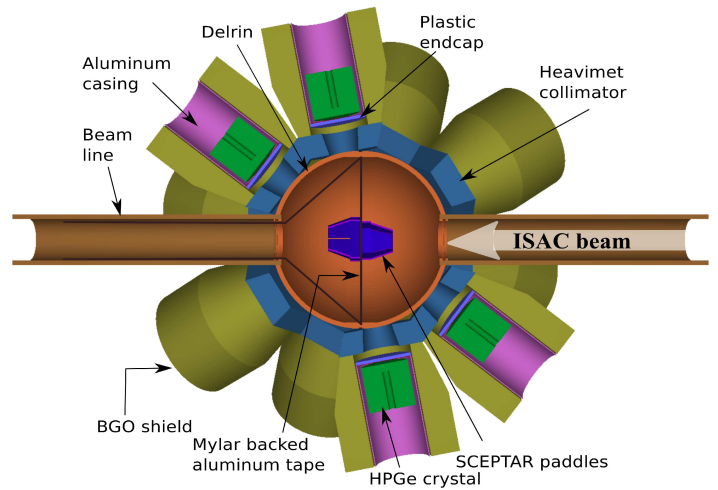}% Here is how to import EPS art
\caption{\label{fig:8pi}Schematic picture of one hemisphere of the $8\pi$ array, shown together with the tape system. This model is drawn to scale and was used for the simulations described in the text.}
\end{figure}

The experiment was carried out using a radioactive $^{19}$Ne ion beam ($T_{1/2} \approx 17.2$~s) at the TRIUMF~Isotope Separator and Accelerator (ISAC) facility located in Vancouver, Canada. The beam was produced %for the measurements
by bombarding a thick, heated SiC target with $\sim 20~\mu$A of 500 MeV protons from the TRIUMF main cyclotron inducing spallation reactions. The diffused reaction products from the
target were then introduced into a forced electron beam-induced arc-discharge (FEBIAD) ion-source via effusion. Subsequently, a pulsed mass-separated beam of \mbox{$\sim 5\times10^5$} $^{19}$Ne ions~s$^{-1}$, with an energy of $\sim$37~keV was delivered to the 8$\pi$ $\gamma$-ray spectrometer~\cite{Garnsworthy:14,Garrett:15}. As shown schematically in Fig.~\ref{fig:8pi}, the spectrometer comprised an array of 20 symmetrically placed Compton-suppressed high-purity germanium (HPGe) detectors, whose inner volume consisted of 20 similarly placed 1.6-mm-thick BC404 plastic scintillator detectors called SCEPTAR~(Scintillating Electron Positron Tagging Array)~\cite{Garnsworthy:14,Garrett:15}. The SCEPTAR detectors were coupled to Hamamatsu H3165-10 photomultiplier tubes (PMTs) and covered $\sim$80\% of the total solid angle. The radioactive ions were implanted on a $\sim$1.3 cm wide, 40~$\mu$m-thick mylar-backed aluminum tape at the center of the 8$\pi$ array. This tape was part of a continuous moving tape 
collector (MTC) 
system that looped in vacuum through a lead-shielded aluminum box located downstream from the array center. The beam pulsing and the MTC allowed for data to be registered using tape cycles.  
In a typical cycle, the beam was implanted on the tape for a certain amount of time, following which the beam was `turned off' using an electrostatic deflector near the ion-source. After a predetermined counting period, the MTC controls were triggered to move any potential long-lived activity on the tape away from the detectors into the shielded tape box. 
%any long-lived radioactive contaminants 
%
\subsection{Data Acquisition}
The data for this experiment were acquired with a fast encoding and read out ADC (FERA) system, with independent data streams for the SCEPTAR and HPGe detectors~\cite{Garrett:15}. The data acquisition (DAQ) trigger included scaled $\beta$ singles and $\beta$-$\gamma$ coincidences for the scintillators and $\gamma$ singles for the HPGe data stream. The events in each data stream were time-stamped using a LeCroy 2367 universal logic module (ULM) acting as a latching scaler, which counted pulses from a precision temperature-stabilized Stanford Research Systems 10~MHz~$\pm$~0.1~Hz oscillator.  

The signals from the 20 SCEPTAR photomultiplier tubes were first shaped by a Phillips Scientific 776 fast amplifier and then sent in parallel to different parts of the data acquisition system. One output from each channel was sent to a LeCroy 4300 charge-to-digital converter (QDC) to generate the minimum ionizing $\beta$ energy spectra. The other output was sent to an Ortec 935 constant fraction discriminator (CFD) for timing purposes. The 500~ns wide pulses from the CFD were also sent in parallel to different units. One branch was converted to 50~ns wide pulses using a fast Phillips Scientific 706 leading edge discriminator and sent to a logic OR fan-in-fan-out unit. The summed pulses from the individual SCEPTAR detectors were imposed with fixed non-extendible dead times in the range of 4-24 $\mu$s, much longer than any processing time in the preceding electronics. The dead-time-affected outputs were finally multiscaled using a Struck SIS3801 multi-channel scaler (MCS). These MCS data were used 
to obtain a high-precision measurement of the $^{19}$Ne half-life, which is described in Ref.~\cite{Triambak:12}. The other branch was sent to multichannel CAEN 894 discriminators, from which the signals were fed to a 32-channel multi-hit LeCroy 3377 time to digital converter (TDC) to store $\beta$ timing information.   

The preamplifier output signals from the HPGe detectors were duplicated as well. The $\gamma$-ray energies were acquired using Ortec 572 spectroscopy amplifiers (with 3~$\mu$s shaping time) and Ortec AD114 peak sensing ADCs. In parallel, the preamplifier signals were sent to Ortec 474 timing filter amplifiers (TFAs) and subsequently discriminated using Ortec 583b CFDs. The fast output of the CFDs were further processed by the TDCs, providing timing information for the $\gamma$-ray events relative to the master trigger signal. These TDCs were additionally used to process the timing from the HPGe bismuth-germanate (BGO) Compton suppression shields, as well as the `inhibit' signals from the pulse pile-up rejection circuitry in the spectroscopy amplifiers. %Ref.~\cite{Garrett:15} describes the $8\pi$ DAQ in greater detail.  
%
% \\
% In addition to the event times for $\beta$ and $\gamma$ ray trigger data, dead times were also recorded on an event-by-event basis. 

Scaled-down $\beta$ singles (with a scale-down factor of 255), $\gamma$ singles and $\beta$-$\gamma$ coincidence data were stored event-by-event in full list mode and reconstructed in an offline analysis.

\section{Analysis}
\subsection{Characteristics of the $\gamma$-ray spectrum}
For this experiment the $8\pi$ MTC was configured~\cite{Triambak:12} so that in each tape cycle we acquired background data for 2~s, following which the $^{19}$Ne ions were collected for $\sim$1-2~s. A counting time of 300~s ($\sim$20 half-lives) was used to collect the decay data. 
\begin{figure}[t]
\includegraphics[scale=0.36]{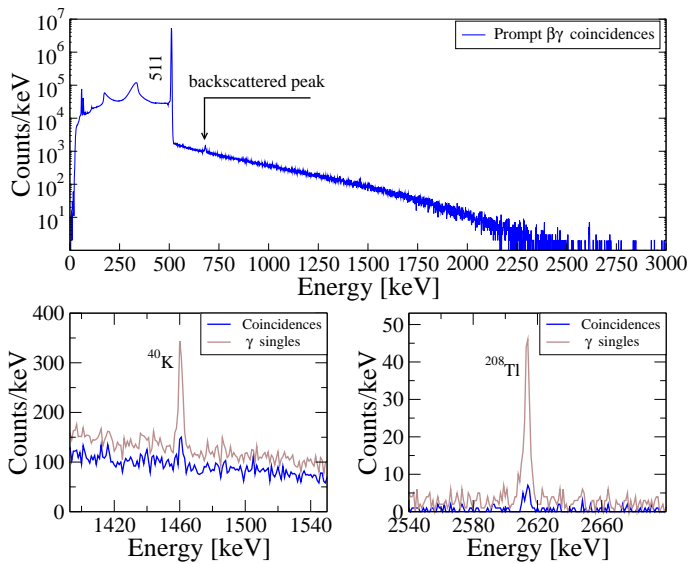}% Here is how to import EPS art
\caption{\label{fig:beta_gamma}Upper panel: $\gamma$-ray spectrum for $^{19}$Ne decay in coincidence with observed positrons. 
%The prompt (black) and random (red) spectra are generated using the gating conditions shown in Fig.~\ref{fig:time_spec}. 
No beam contaminants are apparent from this spectrum, which expectedly is dominated by counts from 511~keV annihilation photons. The broad peak at approximately 680~keV arises from the backscatter of two simultaneous 511~keV $\gamma$ rays. Lower panel: Overlay of $\gamma$~ray singles data with the coincidence spectrum. The data show that room background peaks were significantly reduced by gating on the prompt time differences between successive $\beta$ and $\gamma$ triggers registered with the ULM.  
}
%The contribution from random coincidences is found to be insignificant for this experiment. 
%since several HPGes in the array are positioned at $180^\circ$ to each other
\end{figure}
\begin{figure}[t]
\includegraphics[scale=0.35]{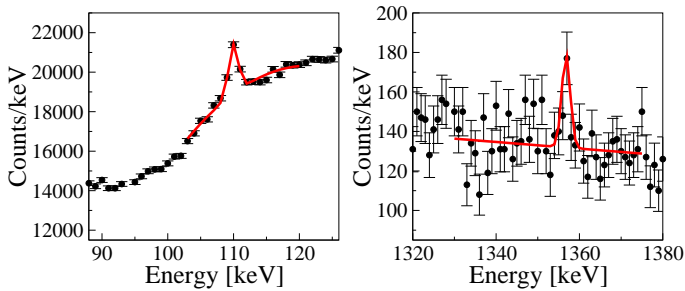}% Here is how to import EPS art
\caption{\label{fig:fits}Fits to the 110~keV and 1357~keV peaks in the coincident $\gamma$-ray spectrum. These peaks arise from the \mbox{$110 \to 0$} and \mbox{$1554 \to 197$~keV} transitions shown in Fig.~\ref{fig:levels}.}
\end{figure}

Figures~\ref{fig:beta_gamma} and \ref{fig:fits} show the $\gamma$-ray spectrum obtained in coincidence with positrons registered with the SCEPTAR detectors. 
Similar to other high precision branching ratio measurements performed with the 8$\pi$ array~\cite{finlay,leach,dunlop} this spectrum was obtained by gating on the time differences between successive $\beta$ and $\gamma$ triggers registered with the ULM. These data were acquired from the same cycle time window that used to determine the total observed $\beta$ singles (which is described in Section~\ref{sec:beta_sing}).  
%We observe that by setting
%a condition on the event times to select only those that were registered within a time difference of 2.4~$\mu$s, as shown in Fig.~\ref{fig:time_spec}, 
%The contribution from randomly correlated \mbox{$\beta$-$\gamma$} events was insignificant for this experiment. 
%This is reflected in the red spectrum shown in Fig.~\ref{fig:beta_gamma}. 
As further illustrated in Fig.~\ref{fig:fits}, we clearly identify two $\gamma$-ray peaks at 110 and 1357~keV that arise from the \mbox{$110 \to 0$} and \mbox{$1554 \to 197$~keV} transitions in the daughter $^{19}$F nucleus. The $\gamma$-ray peak from the \mbox{$197 \to 0$}~keV transition is not visible because of the large Compton artefact in that region of the spectrum. We also do not observe explicit signatures of the (much weaker) 1554 and 1444~keV $\gamma$ rays  (c.f.~Fig.~\ref{fig:levels}) in these data. However, this did not have a bearing in our determination of the $\beta$ decay branches, as discussed below.
\subsection{Efficiency calibration}
\label{sec:eff}
%
% %
% \begin{figure}[t]
% \includegraphics[scale=0.35]{coinc_time_ST.eps}% Here is how to import EPS art
% \caption{\label{fig:time_spec}Time differences between successive $\beta$ and $\gamma$ triggers registered with the ULM. The prompt gate used to generate the spectrum in Fig.~\ref{fig:beta_gamma} is shown in green. We estimate the contribution from randomly correlated $\beta$$\gamma$ events from a background gate of the same width that is shown in red.}
% \end{figure}
% %
%
\begin{figure}[t]
\includegraphics[scale=0.35]{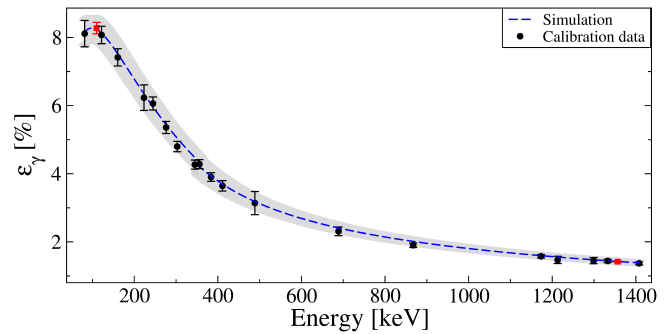}% Here is how to import EPS art
\caption{\label{fig:effi}Experimentally determined $\gamma$-ray efficiencies for the $8\pi$ array in the range $80 \le E_\gamma \le 1408$~keV. The dashed curve outlines normalized simulated efficiencies for individual $\gamma$ rays determined using the PENELOPE code. The gray band represents our conservative estimate of the total uncertainty in the simulations. The red squares show the determined efficiencies at 110 and 1357~keV respectively, that were eventually used to determine the branching ratios of interest.}
\end{figure}
The HPGe detection efficiency for the array was determined using a combination of Monte Carlo simulations performed with the PENELOPE radiation transport code~\cite{penelope} and data obtained from standard $^{133}$Ba, $^{152}$Eu and $^{60}$Co sources. The absolute activities of the latter were known to 3\% at the 99\%CL.
 %Together with , these source data were used to obtain the final $\gamma$ ray efficienies.~%, that used the detector geometry shown in Fig.~\ref{fig:8pi}.. e
The reasons for performing the simulations were two-fold:
\begin{enumerate}
 \item To obtain coincidence summing corrections due to $\gamma$-ray cascades in the calibration sources.
 \item To offer a comparison between the simulated efficiencies and the experimentally determined values.
\end{enumerate}
Figure~\ref{fig:effi} shows the extracted efficiencies for the calibration $\gamma$ rays after applying small corrections due to both coincidence summing as well as pulse pile-up.\footnote{The summing corrections were of the order $\lesssim 2\%$, while the pile-up corrections were of the order $\lesssim 0.3\%$. Incorporating known $\gamma$-$\gamma$ angular correlations in the simulations had an insignificant effect on the former.} These values are found to be in excellent agreement with the results from the PENELOPE simulations (for multiplicity 1 photons), apart from an overall normalization factor. 

Once we ascertained the credibility of the PENELOPE model, the simulations were used both to determine \mbox{$\gamma$-ray} summing corrections for $^{19}$Ne $\beta$ decay (described in Section~\ref{sec:results}) and the attenuation of photons due to absorption in the tape material. %The latter was particularly important for the 110~keV $\gamma$ ray.  
Using a $^{19}$Ne implantation profile from TRIM~\cite{srim,srim2}\footnote{The $^{19}$Ne source was assumed to be uniformly distributed on the tape over a 3~mm radial diameter. TRIM predicts a nearly Gaussian implantation (depth) profile, with a range of $\sim$700~\AA~and straggle of $\sim$290~\AA. The final uncertainties in the extracted efficiencies also included the effect of a (conservative) 1~mm offset in the beam spot laterally.}, the simulations showed that the attenuation was negligible for the 1357~keV $\gamma$ ray, whose efficiency was eventually determined from a polynomial fit to the calibration points
\begin{equation}
ln~\epsilon_\gamma(i) = \sum_{j = 0}^3 a_j~[ln~E_\gamma(i)]^j~.
\end{equation} 
On the other hand, the efficiency for the 110~keV $\gamma$ ray needed a small correction ($\sim 1.5\%$) to the value determined from the above equation, due to $\gamma$-ray absoprtion within the tape. Our extracted efficiencies for both the $\gamma$ rays are highlighted in red in Fig.~\ref{fig:effi}.
% %Consequently, for the final analysis we used the efficiencies 
% %highlighted in red in 
% , that were obtained from 
% 
% %These efficiencies are shown in Fig.~\ref{fig:effi}.
\subsection{$\beta$ singles determination} 
\label{sec:beta_sing}
Similarly as described in Ref.~\cite{iacob:06}, we obtained absolute $\beta$-decay branches to the excited $1/2^-$ and $3/2^+$ states in $^{19}$F from the ratios of $\beta$-$\gamma$ coincident counts to the total number of observed $\beta$ singles. Hence, an important step in our analysis was to obtain the integrated number of $\beta$ singles events ($N_\beta$) detected by the SCEPTAR array. This was determined by a maximum likelihood fit to the total $\beta$ activity (sum of the scaled-up $\beta$ singles and the $\gamma$ coincident $\beta$ decay curves) assuming Poisson-distributed statistics~\cite{Baker:84}. 
The fitted number of counts in each time bin (of width $t_b$) was described by the function
\begin{equation}
\label{eq:dead}
y_{fit}(i) = \dfrac{y(i)}{\left[1+\dfrac{y(i)}{N_{\rm c}t_b}\tau_{\rm eff}\right]}, 
\end{equation}
where
\begin{equation}
\label{eq:expo}
 y(i) = \int_{t_l}^{t_h}A_1 \exp\left({\dfrac{- ln~2}{T_{1/2}}t}\right) dt + \int_{t_l}^{t_h}A_2~dt.
\end{equation}
%comprised an exponentially decaying component on a flat background. 
%so that for an effective dead-time of $\tau_{\rm eff}$ per beta event, 
Eq.~\eqref{eq:dead} represents a realistic model for $N_c$ cycles of data, that are affected by an instrumentation dead time $\tau_{\rm eff}$ per $\beta$ event. 
\begin{figure}[t]
\includegraphics[scale=0.35]{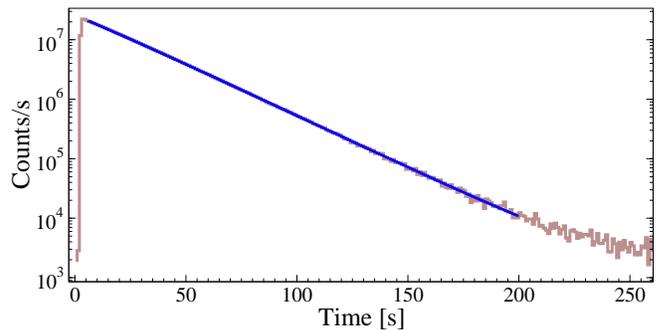}% Here is how to import EPS art
\caption{\label{fig:beta_sing}Best-fit to the dead-time-affected $\beta$ singles activity curve. The total $\chi^2$ for the fit is 30627 for 193 degrees of freedom, which is rather poor. However this has an inconsequential effect on the eventual determination of $\beta$ decay branches. The data described in this analysis were restricted to those runs that were collected at initial $\beta$ rates of $\lesssim$~1~kHz per paddle~\cite{Triambak:12}. In comparison, the measured $\gamma$-ray rates averaged approximately $150$~Hz over the whole experiment. }
\end{figure}
Figure~\ref{fig:beta_sing} shows the summed $\beta$ activity curve generated from the list-mode ULM data, for a total of 724 cycles recorded over the course of our experiment. These data were fitted using fixed values\footnote{The $^{19}$Ne half life used for the fit was from a weighted mean of the results from three recent high-precision measurements~\cite{Triambak:12,Ujic:13,Fontbonne:17}. We refrain from using the result of Broussard \textit{et. al.}~\cite{Broussard:14} due to the inconsistency of their measurement with the other three highest precision results~\cite{Fontbonne:17}.} of $T_{1/2}$ and $\tau_{\rm eff}$, while the $A_1$ and $A_2$ parameters were varied as free parameters. Since the data were the sum of several experimental runs with different implant times, we circumspectly chose the range of the fit to be from $t = 6$~s to $t = 200$~s. This time interval corresponds to approximately 11 half-lives. While it is large enough to provide a reasonably accurate understanding of the background, it 
avoids 
the \mbox{$t > 10T_{1/2}$} region where the $^{19}$Ne activity does not play a statistically significant role.% = 17.257(2)$~s. 

The optimal value of $\tau_{\rm eff}$ was determined from an algorithm that performed the fits described above over a large range of dead times \mbox{$1.5~\mu{\rm s} \le \tau_{\rm eff} \le 4.5~\mu{\rm s}$}, in steps of $\Delta \tau_\text{eff} = 10^{-4}~\mu$s. The $\tau_{\rm eff}$ corresponding to the minimum $\chi^2$ ($\tau_{\rm eff}$ = 3.5~$\mu$s) was eventually used to obtain the total number of $\beta$ singles recorded. 

The best fit to our data on using the optimal value for the effective dead time is shown in Fig.~\ref{fig:beta_sing}. The reduced $\chi^2_{\rm min}$ value for this highly constrained fit is rather poor, but not unexpected, considering the high statistics acquired. 
%Despite the fact that we restricted our analysis only to data sets that were restricted to $\lesssim$ 1~kHz peak rates in each SCEPTAR paddle~\cite{Triambak:12}, 
More realistically, to obtain an improved fit to the data, one would require the incorporation of rate-dependent effects and an accurate understanding of additional complications such as scintillator afterpulsing~\cite{Triambak:12}.  To bypass this problem we choose to make an overly conservative estimation of the total uncertainty on the extracted $N_{\beta}$ value. This was done by using a 99.9\% confidence level uncertainty on $\tau_{\rm eff}$ and further inflating the uncertainty in $A_1$ by a scale factor of $\sqrt{\chi^2/\nu}$ for $\nu$ degrees of freedom. Both these uncertainties were added in quadrature to the uncertainty contribution from the half life value. This procedure yielded a final value of $N_\beta = 5.655(3) \times 10^8$ registered in the time range of the fit.

Although the $\gamma$ ray spectrum in Fig.~\ref{fig:beta_gamma}, and the half-life analysis described in Ref.~\cite{Triambak:12} show no obvious indication of contaminants in the beam, a small contamination of molecular $^{18}{\rm F}^{1}{\rm H}$  cannot be ruled out. However any such contamination would not consequentially affect our measurements, mainly because the long half-life of $^{18}$F~($T_{1/2} \approx 110$~min) would result in an almost constant decay rate over 300~s. Furthermore, the decay of $^{18}$F does not feed any excited levels in $^{18}$O. Therefore its presence in the beam would only result in an increased background component $A_2$ and not affect our $\beta$ singles determination described above.

%to obtain  
\section{Results}
\label{sec:results}
% Below we describe the procedure used to extract the (first-forbidden) $\beta$ branch to the $1/2^-$ excited state in $^{19}$F at 110 keV. A similar procedure was used to obtain the branch to the $3/2^+$ state at 1554~keV. 
If one neglects the small electron-capture fraction for the $\beta$ decay, the ratio of the number of $\beta$-$\gamma$ coincidences for a given $\gamma$ transition from level $i \to j$ to the {\it total} number of $N_\beta$ singles is simply
\begin{equation}
\label{eq:branches}
\dfrac{N_{ij}^{\beta\gamma}}{N_\beta} = \dfrac{1}{\sum_{m}B_m\eta_m}\left[B_i \eta_{i} + \sum_{k>i} B_k \eta_k \gamma_{ki}\right]\gamma_{ij}\epsilon_{ij}~,
\end{equation}
where $B_i$ is the $\beta$ branch to the $i^\text{th}$ level, $\eta_i$ is the $\beta$ detection efficiency for $\beta$ decays feeding level $i$, $\gamma_{ij}$ is the probability of a $\gamma$ transition from level $i$ to $j$ and   $\epsilon_{i j}$ is the efficiency of detecting that $\gamma$ ray. This expression can be simplified further to obtain the $\beta$ branches in Fig.~\ref{fig:levels}. For example, we determine $B_1$ using the photopeak area of the 110~keV $\gamma$ ray and the simple expression
\begin{equation}
B_1 \simeq  k_1\left(\dfrac{N_{10}^{\beta\gamma}}{N_{\beta} \cdot \epsilon_{10}}\right),
\end{equation}
which neglects the contribution with the vanishingly small product \mbox{$B_3\gamma_{31}\epsilon_{10}$}. %\approx 2 \times 10^{-6}$}. 
The $B_3$ branch was obtained similarly. For both these cases, the $N^{\beta\gamma}$ were extracted from a $\gamma$-ray spectrum (shown in Figs.~\ref{fig:beta_gamma} and \ref{fig:fits}) that was projected out from the same cycle time window that was used to determine the total number of $\beta$ singles. 

In the above, $k_1$ is a correction factor \mbox{($\sim 1$)} which accounts for small systematic effects and is imperative for an accurate result. Analogous to the approach followed in Ref.~\cite{iacob:06}, we determined this factor\footnote{Similarly, a correction factor $k_3$ is used to determine $B_3$.} from the product of five distinct corrections that are listed in Table~\ref{tab:errors} and described below.
\begin{table}[t]
\begin{flushleft}
\caption{Relative uncertainties contributing to the first forbidden branch in $^{19}$Ne $\beta^+$ decay.}
\label{tab:errors}
\begin{ruledtabular}
\begin{tabular}{ll.}
Source & \multicolumn{1}{l}{Correction} & \multicolumn{1}{r}{$\frac{\Delta B_1}{B_1}$ (\%)}\\
\colrule
Coincidence summing & 1.0089(6)&0.06\\
Random coincidences & 0.961(9)&0.94\\
Pile up &1.00324(1) &0.001\\
%Dead time &1.00577(6) &\multicolumn{1}{c}{$6\times10^{-3}$}\\
Dead time &1.00577(6) &0.006\\
$Q_\beta$ value dependence on $\beta$ efficiency&1.000(2)
 &0.20\\
$N_{10}^{\beta\gamma}$/$N_\beta$ ratio & &6.4\\
% $\beta$-$\gamma$ coincidences ()& &\\
HPGe efficiency ($\epsilon_{10}$) & &2.4\\
\end{tabular}
\end{ruledtabular}
\end{flushleft}
\end{table}
% \begin{equation}
%  \dfrac{N_{10}^{\beta\gamma}}{N_\beta} \simeq k\left(\dfrac{B_1 \eta_1 \gamma_{10}\epsilon_{10}}{\sum_{i}B_i\eta_i}\right)
% \end{equation}
% \begin{equation}
% \frac{N_{\beta\gamma_{10}}}{N_\beta}= \frac{1}{\sum_i B_i \eta_i}\left[(B_1 \eta_1 \gamma_{10}\epsilon_{10})+(B_3 \eta_3 \gamma_{31}\gamma_{10}\epsilon_{10})\right],
% \end{equation}

 \textit{Summing corrections $(k_s)$ and random coincidences $(k_r)$:} The $\gamma$-ray spectrum in Fig.~\ref{fig:beta_gamma} does not show an explicit signature of photopeak summing with 511~keV annihilation photons, due to the large continuum in the region around 621~keV. Nevertheless, it was important to estimate the photopeak summing with 511~keV $\gamma$~rays, in addition to other summing contributions from scattered positrons, bremsstrahlung and Compton-scattered radiation. Therefore an important part of our analysis was to estimate the coincidence summing corrections $k_s$ for the two $\gamma$ rays of interest. We quantified these corrections with additional PENELOPE Monte Carlo simulations 
 %for 300 million $\beta^+$ decays 
 %that incorporated both the $\beta$ branches and $\gamma$ ray transitions shown in Fig.~\ref{fig:levels}. The simulation 
 %used nominal values of transition probabilities from the literature~\cite{endsf} and 
 that tracked both the positrons and the photons in the active volume of the array, while also taking into account positron annihilation in flight. 
 %\footnote{The simulation used nominal values of transition probabilities from the literature~\cite{endsf} and tracked both the positrons and the photons in the active volume of the array. It also accounted for positron annihilation in flight. Our conservative estimates of the uncertainties in the detector geometry and the branching fractions had a negligible effect on the summing corrections extracted from the simulations.} 
 Our simulations show that roughly 0.9\% of the 110~keV $\gamma$ rays were lost due to coincidence summing. In comparison, the correction for the 1357~keV peak was $k_s = 1.0119(5)$. This value is slightly larger than that for the 110~keV \mbox{$\gamma$-ray} due to an additional contribution from the $1357 \to 197 \to 0$~keV cascade, which is significant and therefore cannot be ignored.

 On the other hand, we determine the correction factor for random $\beta$-$\gamma$ coincidences to be $k_r = 0.961(9)$. This was obtained from the intensity ratios of the background $\gamma$ ray lines observed in the prompt-coincidence and singles $\gamma$ ray spectra (shown in the lower panels of Fig.~\ref{fig:beta_gamma}) together with the known absolute efficiency of the SCEPTAR array.
%obtained from the simulation is shown in Table~\ref{tab:errors}, which  is   
 %The correction factors were obtained from the ratios of efficiencies extracted from this simulation to those obtained in Section~\ref{sec:eff} from more simplistic simulations that used 
 % 
 %The correction factors for the 110 and 1357 keV $\gamma$ rays were obtained from the ratios of revised photopeak efficiencies obtained from these simulations to those obtained in Sectionosts~\ref{sec:eff}. 
 
 \textit{Dead time $(k_d)$ and pile up $(k_p)$:} The latching scalers in the ULM enabled HPGe and SCEPTAR dead times to be calculated independently, on an event-by-event basis~\cite{geoff:nim}. While the SCEPTAR dead time effectively cancels out in Eq.~\eqref{eq:branches}, the $\gamma$-ray photopeak areas required an additional dead time correction. The average dead time per event for the HPGe data stream was found to be 30.4(3)~$\mu$s. Using this value we obtain a HPGe dead time correction factor $k_d = 1.00577(6)$. Independently, we also obtain a pile-up correction $k_p = 1.00324(1)$ from the events registered by the pile-up TDC, that were vetoed from the final $\gamma$-ray spectrum.  
 
 \textit{$\beta$ endpoint energy dependence on SCEPTAR efficiency $(k_\beta)$:} This small correction factor is important for the $B_3$ branch, where the $\beta$ energy distribution is very different than the ones feeding the ground and first excited states in $^{19}$F. It is given by 
 \begin{equation}
  %k_{\beta 3}  = \frac{B_0 \eta_0 + B_1 \eta_1 + B_3 \eta_3}{\eta_3}.
  %k_{\beta 3}  = \frac{1}{\eta_3}\sum_{m=0}^1B_m \eta_m + B_3.
  k_{\beta 3}  = \frac{1}{\eta_3}\sum_{m=0,1,3}B_m \eta_m.
 \end{equation}
We determined this correction from simulations of SCEPTAR efficiencies for the different $Q_\beta$ values feeding the three states of interest at 0, 110 and 1554~keV. The simulations show that the $B_3$ branch requires a correction factor of $k_{\beta 3} = 1.044(2)$. Expectedly this correction for the $B_1$ branch agrees with unity (as $\eta_0/\eta_1 \approx 1$). More detailed investigations of the $Q_\beta$ value dependence on SCEPTAR efficiency can be found in Ref.~\cite{finlay_thesis,*finlay2}.   
% % 
\\
\\
Table~\ref{tab:results} compares our results from this experiment with previous work. While in excellent agreement with earlier measurements\footnote{We do not include the 1975 result of $B_3$ by Freedman {\it et al.}~\cite{Freedman:75} as it significantly disagrees with all subsequent work, including ours.}, our result for the $1/2^+ \to 1/2^-$ first-forbidden branch is $\approx$ 2.4~times more precise than the previous highest-precision measurement. A weighted mean of the results yields final branching ratios of $B_1 = 0.0101(7) \%$ and $B_3 =0.0021(2)\%$. This directly translates to a ground state superallowed branch of 99.9878(7)\%, which is three times more precise than the value reported in a previous compilation~\cite{nathal_prc}.  
\begin{table}[t]
\begin{flushleft}
\caption{A comparison of branching fractions obtained from this measurement to previous work.}
\label{tab:results}
\begin{ruledtabular}
\begin{tabular}{llll}
 & \multicolumn{3}{c}{Measured $\beta$ branch (\%)}\\
\multicolumn{1}{c}{Transition} & \multicolumn{2}{c}{Previous work} & \multicolumn{1}{c}{This work}\\
\colrule
$1/2^+ \to 3/2^+$ & 0.0021(3)$^a$ & 0.0023(3)$^b$&0.0017(5)\\
$1/2^+ \to 1/2^-$ & 0.012(2)$^c$ & 0.011(9)$^d$&0.0099(7)\\
%\colrule
\end{tabular}
\end{ruledtabular}
$^a$ D.\,E.~Alburger~\cite{Alb:76}.\\
$^b$ E.\,G.~Adelberger {\it et al.}~\cite{ega:83}.\\
$^c$ E.\,G.~Adelberger {\it et al.}~\cite{Adel:81}.\\
$^d$ E.\,R.\,J.~Saettler {\it et al.}~\cite{Saett:93}.
\end{flushleft}
\end{table}
Since our published $^{19}$Ne half-life result~\cite{Triambak:12}, there have been three additional half-life measurements reported with comparable or better precision. Similar to our experiment, the authors of Refs.~\cite{Fontbonne:17,Ujic:13} used the method of $\beta$ counting, while Broussard~{\it et~al.}~\cite{Broussard:14} determined the half-life using 511~keV annihilation radiation detected in two collinear HPGe detectors. A weighted mean of the four values yields a poor $\chi^2$ probability of $P(\chi^2,\nu) \approx 1\%$. This is not unexpected, since the $\gamma$-ray measurement disagrees with the other three measurements and is more than $3\sigma$ away than the latest (and most precise) value published in Ref.~\cite{Fontbonne:17}. The probability improves to 65\% if we exclude the value from Ref.~\cite{Broussard:14}. Since this discrepancy is yet to be resolved, for our subsequent analysis we choose to use an average value of $T_{1/2} = 17.257(2)$~s, obtained using only the results from 
Refs.~\cite{Triambak:12,Ujic:13,Fontbonne:17}\footnote{If we include the result from Ref.~\cite{Broussard:14}, the weighted mean changes insignificantly to $T_{1/2} = 17.258(2)$~s.}. 

Together with the electron-capture branching fraction~\cite{toi,bamb,nathal_prc}, the mass excesses from the most recent Atomic Mass Data Center compilation~\cite{amdcweb,amdc:16} and other small corrections~\cite{nathal_prc} due to isospin symmetry breaking and radiative effects, we obtain a corrected $f_Vt$ value for the $1/2^+ \to 1/2^+$ $^{19}$Ne superallowed $\beta$ decay to be
% \begin{equation}
% {\cal F}t = f_V t (1 + \delta_R')(1 + \delta_{NS}^V - \delta_C^V) 
% \end{equation}
\begin{align}
\label{eq:ft}
{\cal F}t^{^{19}{\rm Ne}} & = f_V t (1 + \delta_R')(1 + \delta_{NS}^V - \delta_C^V)\nonumber\\
& = 1721.44(92)~{\rm s},
\end{align}
where we follow the same notation as Refs.~\cite{nathal_prc,TH:15} and $f_V = 98.649(31)$ is the vector component of the statistical rate function for the transition.\footnote{This is slightly different than the axial-vector part, mainly because of the effect of weak magnetism~\cite{Triambak:17,Grenacs} on the shape-correction factor of the latter~\cite{iacob:06}. We obtain $f_A/f_V = 1.0142(28)$~\cite{ian_pvt}, where, similar to Ref.~\cite{oscar:09}, we assign a 20\% relative uncertainty on the deviation of $f_A/f_V$ from unity.}     

As a result of the aforementioned high-precision half-life and $\beta$ branch measurements, the value in Eq.~\eqref{eq:ft} is now one of the most precisely measured ${\cal F}t$ values for \mbox{$T = 1/2 \to T = 1/2$} mirror transitions. Consequently, it provides a benchmark for comparison with experimental observables that are used for searches of BSM physics. We discuss some examples below. 
   
%
%\subsection{The $^{19}$Ne half-life}
%\subsubsection{Systematic effects}
%\subsectiowhn{Beta branches}
%\subsubsection{Systematic effects}

\section{Discussion}
\subsection{Implications for searches of second-class weak interactions}
\label{secondclass}
Beyond the allowed approximation, the hadronic weak current contains additional {\it recoil-order} form factors~\cite{Grenacs,Triambak:17,Holstein:rmp,crigliano}.  %together with the usual vector and axial-vector terms. 
Some of these terms are classified as \textit{second-class}, based on their transformation properties under the $G$-parity operation~\cite{Weinberg,Triambak:17}. Within the limit of perfect isospin symmetry, second-class currents are forbidden in the standard model. Angular correlation measurements in nuclear $\beta$ decays are known to be useful probes to search for induced second-class currents~\cite{Holstein:rmp,Holstein:71}. As an example, we focus on the $\beta^+$ decay of spin-polarized $^{19}$Ne nuclei. After integrating over the neutrino directions, the differential decay rate can be expressed in terms of the spectral functions $f_i(E)$~\cite{Holstein:71} 
%(after integrating over the neutrino directions), 
\begin{equation}
\label{eq:rate}
d\Gamma \propto (E_0 - E)^2p E \left\{f_1(E) + f_4(E)\frac{\langle \bm{J}\rangle}{J}\cdot \frac{\bm{p}}{E}+...\right\}dE d\Omega_e~,
\end{equation}
where 
\begin{widetext}
\begin{equation}
\label{eq:f1}
f_1(E) = a^2 + c^2 -\frac{2E_0}{3 M} (c^2 -bc -cd) +\frac{2E}{3M}(3a^2 + 5c^2 -2bc)-\frac{m_e^2}{3ME}(2c^2-2bc-cd), \\
\end{equation}
and
\begin{align}
\label{eq:f4}
f_4(E) & =\sqrt{\frac{J}{J+1}}\left[2ac - \frac{2E_0}{3M}(ac-ab-ad) + \frac{2E}{3M}(7ac -ab -ad)\right]\nonumber\\
&+\left(\frac{1}{J+1}\right)\left[c^2 - \frac{2E_0}{3M}(c^2 -bc -cd)+ \frac{E}{3M}(11c^2 - 5bc + cd) \right].   
 \end{align}
\end{widetext}
In the above, $J = 1/2$, $E$ is the total energy of the positrons, $E_0$ is the end-point energy, $\bm{p}$ is the positron momentum, $m_e$ is the positron mass and $M$ is the average of the parent and daughter masses.  The remaining terms are momentum-transfer dependent form factors; $a(q^2)$ and $c(q^2)$ are the leading vector and axial-vector form factors, $b(q^2)$ is the weak magnetism form factor and $d(q^2)$ is an induced-tensor form factor. It is apparent from Eq.\eqref{eq:rate} that if one ignores small electromagnetic corrections due to final-state Coulomb interactions~\cite{holstein:coul}, then the $\beta$ asymmetry parameter $A_\beta(E)$ for the decay can be defined in terms of these spectral functions, so that \mbox{$A_\beta= f_4(E)/f_1(E)$}. 

In the low-momentum transfer ($q^2 \to 0$) limit, \mbox{$a = C_V M_F$} and \mbox{$c = C_A M_{GT}$}, where $M_F$ and $M_{GT}$ are the usual Fermi and Gamow-Teller matrix elements~\cite{nathal_prc}. Both these and the other energy dependent (recoil-order) terms in the spectral functions can be determined using the conserved vector current (CVC) hypothesis~\cite{cvc1}. For $^{19}$Ne $\beta^+$ decay, the vector and weak magnetism form factors reduce to $a = 1$ and $b = -148.5605(26)$~\cite{oscar:09}, where the latter is calculated from the magnetic moments of the parent and daughter nuclei~\cite{Holstein:71}. The standard-model-allowed (first-class) contribution to the induced-tensor form factor is expected to be highly suppressed as the decay mainly occurs between isobaric analog states~\cite{Holstein:71}. Finally, the standard model value for the Gamow-Teller form factor $c$ can be extracted 
from the averaged ${\cal F}t$ value of \mbox{$0^+ \to 0^+$} superallowed Fermi transitions~\cite{TH:15} (or equivalently $V_{ud}$) and ${\cal F}t^{^{19}{\rm Ne}}$. We determine this to be \mbox{$c_{\rm SM} = -1.5916(23)$}.\footnote{This form factor has a negative sign because we follow the same representation for Dirac matrices as Ref.~\cite{Holstein:71}. Standard-model-allowed recoil-order corrections~\cite{Holstein:rmp} are taken into consideration in this calculation and hereafter.}
% 
% , . This value is eventually used to determine expected angular correlation coefficients (such as $A_\beta$), assuming only the minimal standard model. 
%
\begin{table}[t]
\begin{flushleft}
\caption{Measured $A_\beta$ values (in \%) for $^{19}$Ne superallowed decay. For comparison we list the standard model predictions obtained using the ${\cal F}t$ value in Eq.~\eqref{eq:ft}.}
\label{tab:comp_beta}
\begin{ruledtabular}
\begin{tabular}{lc.c}
Year &\multicolumn{1}{c}{Reference} &\multicolumn{1}{c}{$A_{\beta}(0)$$^a$}& \multicolumn{1}{c}{$A_\beta$$^b$}\\
 \colrule
1963&Commins and Dobson~\cite{Commins:63}&...&$-5.7(5)$\\
1967&Calaprice~{\it et al.}~\cite{Calaprice:67}&...&$-3.3(2)$\\
1969&Calaprice~{\it et al.}~\cite{Calaprice:69}&...&$-3.9(2)$\\
1975&Calaprice~{\it et al.}~\cite{calaprice:75}&-3.91(14)&...\\
1983&Schreiber~\cite{Schreiber}&-3.603(83)&...\\
1996&Jones~\cite{Jones}&-3.52(11)&...\\
%\colrule
%\multicolumn{2}{c}{Weighted mean} &-3.65(6)&$-3.76(44)^c$
\end{tabular}
\end{ruledtabular}
$^a$ Standard model prediction for $A_{\beta}(0) = -4.15(6)$\%.\\
$^b$ Standard model prediction for $A_\beta = -4.49(6)$\%.\\
%$^c$ Since these measurements are in conflict with one another we inflated the uncertainty in the mean by a scale factor $\sqrt{\chi^2/2} = 3.22$.
 \end{flushleft}
 \end{table}

 \begin{table}[t]
\begin{flushleft}
\caption{Measured $dA_\beta/E$ values for $^{19}$Ne superallowed decay. Similarly as in Table~\ref{tab:comp_beta}, the standard model prediction for the slope is listed for comparison.}
\label{tab:comp_slope}
\begin{ruledtabular}
\begin{tabular}{ll.}
Year &Reference &\multicolumn{1}{c}{$dA_{\beta }/dE$$^a$}\\
     &          &\multicolumn{1}{c}{(\% MeV$^{-1}$)}\\
 \colrule
1975&Calaprice~{\it et al.}~\cite{calaprice:75}&-0.65(15)\\
1983&Schreiber~\cite{Schreiber}&-0.486(77)\\
1996&Jones~\cite{Jones}&-0.42(11)\\
%\colrule
%\multicolumn{2}{c}{Weighted mean} &-0.492(58)
\end{tabular}
\end{ruledtabular}
$^a$ $(dA_{\beta}/dE)_{\rm SM} = -0.349(2)$\%~MeV$^{-1}$.\\
 \end{flushleft}
 \end{table}
\begin{figure}[t]
\includegraphics[scale=0.35]{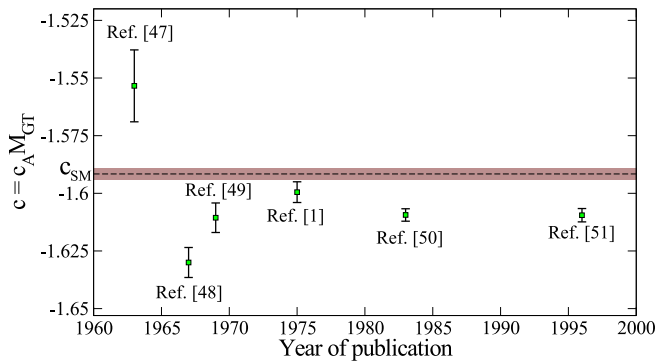}% Here is how to import EPS art
\caption{\label{fig:gt_matrix}The axial-vector form factor for $^{19}$Ne $\beta^+$ decay obtained from independent beta asymmetry measurements. The standard model prediction $c_{\rm SM}$ obtained from this work is shown for comparison.}
\end{figure}
 There have been several measurements of $^{19}$Ne $\beta^+$ decay asymmetry performed in the past. These are listed in Table~\ref{tab:comp_beta}.
 Furthermore, since the weak magnetism and induced-tensor form factors affect the energy dependence of $A_\beta$, a measurement of the slope $dA_\beta/dE$ allows a sensitive search for second-class currents. This approach was first used by Calaprice~\textit{et al.}~\cite{calaprice:75} to search for second-class currents in $^{19}$Ne decay. 
%They obtain $(dA_\beta/dE)_{\rm exp} = -0.65(15)\%$, which requires 
Interestingly, their measurement significantly disagreed with CVC predictions, requiring 
an unexpectedly large second-class tensor form factor to describe the data. Calaprice~\textit{et~al.} also reported a zero kinetic energy intercept value for the beta asymmetry, which is listed as $A_\beta(0)$ in Table~\ref{tab:comp_beta}. This work was followed by two other measurements whose results were never published, but reported in Ph.D. theses~\cite{Schreiber,Jones}. Although the $dA_\beta/dE$ results from the three experiments are in reasonable agreement with each other, the unpublished values are more consistent with the standard model prediction and other experimental results that do not show explicit signatures of second class currents~\cite{tribble1,tribble2,kei1,kei2}.  

For completeness we list these results together with earlier $\beta$ asymmetry measurements in Tables~\ref{tab:comp_beta} and~\ref{tab:comp_slope}. It is worthwhile to note that unlike \mbox{Refs.~\cite{calaprice:75,Schreiber,Jones}}, the older  measurements~\cite{Commins:63,Calaprice:67,Calaprice:69} were performed by integrating over the whole positron spectrum.

If one assumes $d = 0$, the measured $A_\beta$ coefficient can be used to determine the axial-vector form factor for the decay, independent of other standard model expectations. This is shown in Fig.~\ref{fig:gt_matrix}, where we plot the value for $c$ extracted from all previous $\beta$ asymmetry measurements\footnote{For the $A_\beta(0)$ measurements $E = 0.511$~MeV. For the others we use an averaged value of $E = 1.474$~MeV for the positrons.} for $^{19}$Ne decay. Clearly, these data are in conflict with the CVC prediction and with each other. Some consequences of these differences are discussed below. 
%which warrants further investigation. This is further elucidated below.      
% 
%They are clearly in conflict with each other. Consequently we had to multiply the weighted uncertainty from these results by scale factor $S = \sqrt{\chi^2/2}$. 

\subsection{Implications for searches of right-handed currents}
Despite the observed $V-A$ character of weak interactions, some of the earliest extensions to the standard model~\cite{pati:74,senjanovic:75} and their more modern versions~\cite{Shaban,Herczeg} use a parity symmetric  Lagrangian~\cite{beg:77} to describe the theory. These models restore parity at a higher energy scale,  and provide a framework within which the apparent non-conservation of parity at lower energies can be attributed to the spontaneous breakdown of a higher gauge symmetry~\cite{Herczeg,beg:77,Langacker:89}. The extended gauge group requires the existence of additional right-handed $W$ and $Z$ bosons, which are much heavier than their left-handed counterparts. 

Such models present a compelling case. Not only does the inherent left-right (LR) symmetry make them aesthetically pleasing, the suppression of $V+A$ type weak interactions at low energies is a natural consequence in these models, owing to the large masses of the right-handed gauge bosons. It has also been shown that this 
suppression has a direct relation to both the smallness of neutrino masses~\cite{mohapatra:80} as well as the experimentally observed $CP$ violation~\cite{Branco}. %This makes left-right (LR) symmetric models even more appealing.

In the simplest (manifest) LR models~\cite{beg:77}, the left-handed and right-handed charged weak currents couple to the weak interaction eigenstates $W_L$ and $W_R$ and have identical transformation  properties (apart from chirality).\footnote{Here, the left-handed and right-handed sectors have identical coupling constants and mixing angles. There are no additional $CP$ violating phases apart from the usual Kobayashi-Maskawa phase~\cite{Herczeg}.} On account of the symmetry breaking, the mass eigenstates are simply linear combinations of the weak interaction eigenstates, with a LR mixing angle $\zeta$~\cite{beg:77}. The weak interaction can therefore be parameterized~\cite{Carnoy:PRD} in terms of $\zeta$ and the ratio $\delta = (M_1/M_2)^2$, where $M_1~(M_2)$ is the mass of the left (right) handed boson, with $M_1 \ll M_2$. Following Holstein and Treiman~\cite{holstein:77} and B\'eg~{\it et al.}~\cite{beg:77}, one can further define two new parameters $x$ and $y$, which are related to $\delta$ and $\zeta$. 
For sufficiently small $\delta$ and $\zeta$, these reduce to $x \simeq \delta - \zeta$ and $y \simeq \delta + \zeta$~\cite{Carnoy:PRD}. Such a prescription ensures that purely left-handed weak interactions would emerge for vanishing values of $x$ and $y$. 

The above parameterization modifies the $f_1$ and $f_4$ spectral functions to allow for right handed currents (RHCs), such that~\cite{holstein:77}
\begin{align}
 f_1 & \to f_1 + x^2a^2 + y^2 c^2 \\
\label{eq:mod_fs}
 f_4 &\to f_4 - \frac{y^2 c^2}{J+1} - 2\sqrt{\frac{J}{J+1}}xyac. 
\end{align}
This makes the experimentally measured $A_\beta$ parameter sensitive to right-handed weak interactions.\footnote{This analysis is valid only if the RH neutrinos are light enough not to kinematically suppress the decay.} 

We have already shown in Table.~\ref{tab:comp_beta} that the $\beta$ asymmetry for $^{19}$Ne decay is quite small. This is due to an accidental cancellation of the leading form factors in Eq.~\eqref{eq:f4}. Evidently, a similar cancellation does not take place for the RHC contribution in Eq.~\eqref{eq:mod_fs}, except when $x = y$.  This makes $^{19}$Ne $\beta$ decay highly sensitive to RHCs. As a matter of fact, it is the most sensitive probe for RHCs among all mirror transitions up to \mbox{$A = 41$}~\cite{oscar:91}. For example, using a `sensitivity coefficient' defined by the authors of Ref.~\cite{oscar:91}, it is calculated to be $\sim$~70 times more sensitive~\cite{oscar:91} than $^{37}$K $\beta$ decay, whose beta asymmetry was recently reported~\cite{fenker:18} with the highest relative precision amongst all $T = 1/2$ nuclides.

Unless the ratio of axial-vector and vector form factors for the decay is determined independently (e.g. from a \mbox{$\beta$-$\nu$} correlation measurement), a stand alone $\beta$ asymmetry measurement by itself cannot be used to place constraints on allowed values of $\delta$ and $\zeta$. In facing such a scenario for $^{19}$Ne $\beta$ decay, one has to resort to the ratio
\begin{equation}
 R = \left(\dfrac{{\cal F}t^{0^+ \to 0^+}}{{\cal F}t^{^{19}{\rm Ne}}}\right), 
\end{equation}
where ${\cal F}t^{0^+ \to 0^+}= 3072.27(72)$~s is the averaged ${\cal F}t$ value from \mbox{$0^+ \to 0^+$} superallowed Fermi transitions~\cite{TH:15}.

If one permits the existence of RHCs $(x,y \neq 0)$, then $R$ can be expressed as~\cite{holstein:77}
\begin{equation}
%R =  \frac{1}{2}\left[1+\frac{f_A}{f_V}\rho^2\frac{1+y^2}{1+x^2}\right]
R \simeq \left[\frac{a^2(1+x^2) + \left(\frac{f_A}{f_V}\right)c^2(1+y^2)+ r_i}{2 a^2 (1 + x^2)}\right],
\end{equation}
where $r_i$ are the small recoil-order corrections in Eq.~\eqref{eq:f1}. Therefore, it is imperative that \textit{both} $R$ and $A_\beta$ are known with high precision and accuracy, in order to place meaningful bounds on RHCs. On using the currently determined high-precision value for ${\cal F}t^{^{19}{\rm Ne}}$, we obtain $R = 1.785(1)$, which is three times more precise than known previously.
\begin{figure}[t]
\includegraphics[scale=0.36]{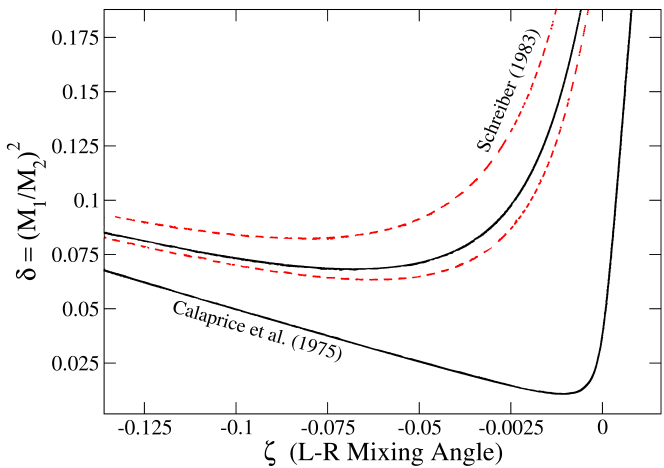}% Here is how to import EPS art
\caption{\label{fig:contour}A comparison of 90\% CL constraints on $\delta$ and $\zeta$ set by two independent $A_\beta(0)$ measurements for $^{19}$Ne decay (black solid line~\cite{calaprice:75} and red dashed line~\cite{Schreiber}), together with the ${\cal F}t$ value obtained in this work. The allowed regions were generated assuming the manifest LR model. 
}
\end{figure}

In Fig.~\ref{fig:contour} we show the 90\% CL allowed region in the $(\delta,\zeta)$ parameter space, obtained\footnote{We set $d = 0$ in this part of the analysis and the next subsection.} from a simultaneous fit to $R$ and the beta asymmetry measured by Calaprice~{\it et~al.}~\cite{calaprice:75}. Despite the fact that the measured $dA_\beta/dE$ from the same experiment yielded a much larger slope than expected, we choose this value of $A_\beta(0)$ for the following reasons.  Firstly, it is the latest (and most precise) {\it published} $\beta$ asymmetry measurement for $^{19}$Ne decay. Secondly, together with the ${\cal F}t^{^{19}{\rm Ne}}$ value in Eq.~\eqref{eq:ft}, the other asymmetry measurements of Refs.~\cite{Calaprice:67, Calaprice:69,Schreiber,Jones} yield values for the $V_{ud}$ matrix element (c.f. Section.~\ref{vud}) that are much smaller than expected. Consequently, together with the current Particle Data Group (PDG) recommended values~\cite{pdg1,pdg2} for $V_{us}$ and $V_{ub}$, these results 
lead to significant violations of the CKM unitarity condition.\footnote{The extracted values of $V_{ud}$ from the results of Refs.~\cite{Calaprice:67, Schreiber,Jones} result in a violation of CKM unitarity by 5 standard deviations or more. The value obtained from the 1969 measurement~\cite{Calaprice:69} can also be ruled out as it disagrees with unitarity at the 99.6\%~CL.} For the sake of comparison we also show in Fig.~\ref{fig:contour} the $90\%$ CL bounds obtained using the most precise reported (\textit{unpublished}) value of the $\beta$ asymmetry parameter by Schreiber~\cite{Schreiber}. The ratio of the Gamow-Teller to Fermi form factors obtained from this measurement is in almost exact agreement with the independent determinations of Refs.~\cite{Calaprice:69,Jones} (c.f.~Fig.\ref{fig:gt_matrix}). Furthermore, the energy dependence $dA_\beta/dE$ determined from Schreiber's experiment shows no indication of second-class currents and is also in excellent agreement with the later 
measurement by 
Jones~\cite{Jones}. However, despite this consistency, the quoted $A_\beta(0)$ value from this experiment shows a significantly large signal for RHCs, as apparent in Fig.~\ref{fig:contour}. This should not be surprising, given the CKM unitarity violation mentioned previously. The best fit to these data disagrees with the standard model prediction for no RHCs ($\delta = \zeta = 0$) at the $6.4\sigma$ level. In contrast, the best fit using the result from Ref.~\cite{calaprice:75} deviates from the standard model by only $1.7\sigma$.

In light of the above, we conclude that the systematic effects that might have affected the $A_\beta$ slope measurement in Ref.~\cite{calaprice:75} did not significantly influence their extraction of the zero kinetic energy intercept value $A_\beta(0)$. All the other measured values for the $\beta$ asymmetry (except the lowest precision measurement from 1963) can be ruled out. There has been a recent effort~\cite{combs} to reanalyze the data acquired by Ref.~\cite{Jones}, by placing emphasis on positron backscattering and other systematic effects. The results from this reanalysis are expected to be published soon~\cite{young}.
%It is quite likely that backscattering is the major systematic effect that plaques the accuracy of these challenging measurements.  

\subsection{A determination of {\boldmath $V_{ud}$}}
\label{vud}
It was implicit in the previous discussion that if one assumes conservation of the vector current, the ${\cal F}t$ value in Eq.~\eqref{eq:ft} determines~\cite{oscar:09} the $V_{ud}$ element of the CKM quark-mixing matrix. The expression to obtain $V_{ud}$ is analogous to neutron decay, where 
\begin{equation}
V_{ud} = \left[\frac{K}{{\cal F}t G_F^2}\frac{1}{(1+\Delta_R^V)(1+\frac{f_A}{f_V}\rho^2 + r_i)}\right]^{1/2}.
\end{equation}
Here, $K/(\hbar c)^6 = 2\pi^3 \hbar~ln 2  /(m_e c^2)^5 = 8120.2776(9)\times10^{-10}~{\rm GeV}^{-4}~{\rm s}$~\cite{TH:15}, $G_F/(\hbar c)^3 = 1.1663787(6) \times 10^{-5}~{\rm GeV}^{-2}$ is the universal Fermi coupling constant~\cite{mulan}, \mbox{$\Delta_R^V = 2.361(38)\%$} is a nucleus independent electroweak radiative correction~\cite{marciano:06} and $\rho = c/a$. Needless to say, determining $V_{ud}$ in this manner requires an independent correlation measurement to obtain the mixing ratio $\rho$. We obtain $\rho = -1.5995(45)$ from the $A_\beta(0)$ measurement of Ref.~\cite{calaprice:75}. Using this value of $\rho$ and the ${\cal F}t$ value determined in this work, we obtain $V_{ud} = 0.9707(22)$. This is in reasonable agreement with the high precision value extracted from superallowed $0^+ \to 0^+$ Fermi transitions~\cite{TH:15}.

It should be noted that the radiative correction mentioned above was recently revaluated to be \mbox{$\Delta_R^V = 2.467(22)\%$}~\cite{Seng:2018} using dispersion relations together with neutrino scattering data. However, incorporating this new result has an insignificant effect on our extracted value for $V_{ud}$, as the latter's uncertainty is dominated by the uncertainty contribution from $A_\beta(0)$.  

% 
% In order to accurately extract $\rho$ from these values it is essential to take into account small corrections\footnote{These are slightly different for the two measurements, as the authors of Ref.~\cite{Commins:63} obtained a value integrated over the whole $\beta$ spectrum, while Calaprice {\it et al.} measured $A_\beta$ as a function of positron kinetic energy and reported the zero energy intercept value $A_\beta(0)$.} arising from recoil-order effects~~\cite{Holstein:rmp,Grenacs,holstein:77}.      

%, particularly the weak magnetism form factor~\cite{Triambak:17,Grenacs}. 
%
\subsection{Towards a better understanding of parity violating NN interactions}
The $J^\pi = (1/2^+;~1/2^-$), $T = 1/2$ parity doublet in $^{19}$F (shown in Fig.~\ref{fig:levels}) plays an important role in elucidating both the isoscalar and isovector components of $\Delta S = 0$ parity non-conserving (PNC) hadronic weak interactions~\cite{Adelberger:85,haeberli}. It is one of the few cases where the PNC NN interaction admixes the doublet states significantly, on account of the small energy spacing between the levels ($\Delta E = 110$~keV) and the absence of other nearby $J = 1/2$ states. This leads to an amplification of the parity violating observable, namely the asymmetry of the 110~keV $\gamma$ rays that would be emitted from a polarized ensemble of $^{19}$F nuclei in the first excited $1/2^-$ state. The $\gamma$ asymmetry has been measured independently by two groups, whose results are in excellent agreement with each other~\cite{ega:83,elsener}. However, these experimental results are approximately three times smaller than shell model predictions~\cite{haxton:80,brown:80} that used the `best values' recommended 
by Desplanques, Donoghue and Holstein~\cite{Adelberger:85,haeberli,DDH} for the weakly interacting PNC meson-nucleon couplings.

It has been shown that the model dependence in extracting the weak NN amplitudes from the PNC observables can be largely minimized in such cases via measurements of the $\beta$ decay transition rates connecting the isobaric analog of one member of the doublet to the other~\cite{ega:83}. Here it is the first-forbidden $1/2^+ \to 1/2^-$ transition in $^{19}$Ne $\beta^+$ decay. In the $q^2 \to 0$ limit, the forbidden $\beta$ decay matrix element is dominated by the $\Delta J^\pi = 0^-$ axial-charge operator, and is very similar to the isovector part of the parity mixing matrix element.  Therefore, not only does the measured $1/2^+ \to 1/2^-$ $^{19}$Ne $\beta$ decay rate offer a model-independent means to calculate the isovector PNC NN amplitudes, it also allows a useful check of the wavefunctions that are used to analyze the parity mixing in $^{19}$F. A previous analysis showed that the calculated decay rate was about 10~times larger than the measured values~\cite{Adel:81}, presumably because of the omission of 
$5p-
2h$ correlations in the shell model wavefunctions. This would explain the factor of 3 discrepancy between the measured and calculated values of the $\gamma$ asymmetry mentioned previously, assuming that the isoscalar contribution of the parity violating matrix element also scales similarly~\cite{ega:83}. It has been suggested~\cite{Adel:81,ega:83} that a large-basis shell model calculation which includes $2\hbar\omega$ excitations would resolve this issue.     

In light of the above and the recent development of state-of-the-art computational techniques~\cite{wasem,philips,schindler,haxton:13,gardner} to extract elementary PNC amplitudes, we anticipate our high precision measurement of the first-forbidden branch will be useful to constrain future calculations. Together with the high-precision values for the $^{19}$Ne half-life and a weighted mean of the results in Table~\ref{tab:results}, we obtain a first-forbidden transition rate of $\omega_{\rm expt} = 4.06(28)\times10^{-6}~{\rm s}^{-1}$. On further assuming an allowed spectrum shape\footnote{This is a reasonable approximation as the axial-charge operator is independent of the momentum transferred to the leptons~\cite{Adel:81,ega:83}.} for the first forbidden transition~\cite{Adel:81} we determine its $ft$ value to be $1.35(9)\times10^7$~s. Our values are roughly two times more precise, yet in agreement with previous measurements.

% and with shell model calculations~\cite{Adelberger:85, ega:83}.  

%Furthermore, one can obtain reasonably accurate wavefunctions for the doublet states in such light nuclei.  
% % \subsection{Tensor currents}
% \subsection{The first-forbidden {\boldmath $1/2^+ \to 1/2^-$} branch and investigations of the PNC {\boldmath $nn$} interaction}
\section{Conclusions}
We measured $\beta$ decay branches to excited states in $^{19}$F for the first time using a radioactive $^{19}$Ne beam. Unlike previous measurements that used $(p,n)$ reactions on gas targets, our experiment was minimally affected by the source distribution and other associated systematic effects. We obtain high precision values for the $\beta$ transition rates that would be useful for a variety of fundamental symmetry tests that involve $^{19}$Ne and $^{19}$F nuclei.     
\begin{acknowledgments}
We thank Ian~Towner for calculating the statistical rate functions. We are also grateful to Alejandro Garc\'ia, Eric Adelberger, Wick Haxton and Dan Melconian for fruitful discussions.  
%We also benefited greatly from fruitful discussions with Wick Haxton.
This work was partially funded by the National Research Foundation (NRF) of South Africa and the Natural Sciences and Engineering Research Council of Canada (NSERC). PZM thanks the NRF funded MANUS/MATSCI program at UWC/UZ for financial support during the course of her M.Sc. 
TRIUMF receives federal funding via a contribution agreement through the National Research Council of Canada.
\end{acknowledgments}

%\nocite{*}
%\bibliographystyle{apsrev4-1.bst}
\bibliography{19ne_ft}% Produces the bibliography via BibTeX.

\begin{thebibliography}{82}%
\makeatletter
\providecommand \@ifxundefined [1]{%
 \@ifx{#1\undefined}
}%
\providecommand \@ifnum [1]{%
 \ifnum #1\expandafter \@firstoftwo
 \else \expandafter \@secondoftwo
 \fi
}%
\providecommand \@ifx [1]{%
 \ifx #1\expandafter \@firstoftwo
 \else \expandafter \@secondoftwo
 \fi
}%
\providecommand \natexlab [1]{#1}%
\providecommand \enquote  [1]{``#1''}%
\providecommand \bibnamefont  [1]{#1}%
\providecommand \bibfnamefont [1]{#1}%
\providecommand \citenamefont [1]{#1}%
\providecommand \href@noop [0]{\@secondoftwo}%
\providecommand \href [0]{\begingroup \@sanitize@url \@href}%
\providecommand \@href[1]{\@@startlink{#1}\@@href}%
\providecommand \@@href[1]{\endgroup#1\@@endlink}%
\providecommand \@sanitize@url [0]{\catcode `\\12\catcode `\$12\catcode
  `\&12\catcode `\#12\catcode `\^12\catcode `\_12\catcode `\%12\relax}%
\providecommand \@@startlink[1]{}%
\providecommand \@@endlink[0]{}%
\providecommand \url  [0]{\begingroup\@sanitize@url \@url }%
\providecommand \@url [1]{\endgroup\@href {#1}{\urlprefix }}%
\providecommand \urlprefix  [0]{URL }%
\providecommand \Eprint [0]{\href }%
\providecommand \doibase [0]{http://dx.doi.org/}%
\providecommand \selectlanguage [0]{\@gobble}%
\providecommand \bibinfo  [0]{\@secondoftwo}%
\providecommand \bibfield  [0]{\@secondoftwo}%
\providecommand \translation [1]{[#1]}%
\providecommand \BibitemOpen [0]{}%
\providecommand \bibitemStop [0]{}%
\providecommand \bibitemNoStop [0]{.\EOS\space}%
\providecommand \EOS [0]{\spacefactor3000\relax}%
\providecommand \BibitemShut  [1]{\csname bibitem#1\endcsname}%
\let\auto@bib@innerbib\@empty
%</preamble>
\bibitem [{\citenamefont {Calaprice}\ \emph {et~al.}(1975)\citenamefont
  {Calaprice}, \citenamefont {Freedman}, \citenamefont {Mead},\ and\
  \citenamefont {Vantine}}]{calaprice:75}%
  \BibitemOpen
  \bibfield  {author} {\bibinfo {author} {\bibfnamefont {F.~P.}\ \bibnamefont
  {Calaprice}}, \bibinfo {author} {\bibfnamefont {S.~J.}\ \bibnamefont
  {Freedman}}, \bibinfo {author} {\bibfnamefont {W.~C.}\ \bibnamefont {Mead}},
  \ and\ \bibinfo {author} {\bibfnamefont {H.~C.}\ \bibnamefont {Vantine}},\
  }\href {\doibase 10.1103/PhysRevLett.35.1566} {\bibfield  {journal} {\bibinfo
   {journal} {Phys. Rev. Lett.}\ }\textbf {\bibinfo {volume} {35}},\ \bibinfo
  {pages} {1566} (\bibinfo {year} {1975})}\BibitemShut {NoStop}%
\bibitem [{\citenamefont {Carnoy}\ \emph {et~al.}(1992)\citenamefont {Carnoy},
  \citenamefont {Deutsch}, \citenamefont {Prieels}, \citenamefont {Severijns},\
  and\ \citenamefont {Quin}}]{carnoy:92}%
  \BibitemOpen
  \bibfield  {author} {\bibinfo {author} {\bibfnamefont {A.~S.}\ \bibnamefont
  {Carnoy}}, \bibinfo {author} {\bibfnamefont {J.}~\bibnamefont {Deutsch}},
  \bibinfo {author} {\bibfnamefont {R.}~\bibnamefont {Prieels}}, \bibinfo
  {author} {\bibfnamefont {N.}~\bibnamefont {Severijns}}, \ and\ \bibinfo
  {author} {\bibfnamefont {P.~A.}\ \bibnamefont {Quin}},\ }\href
  {http://stacks.iop.org/0954-3899/18/i=5/a=011} {\bibfield  {journal}
  {\bibinfo  {journal} {Journal of Physics G: Nuclear and Particle Physics}\
  }\textbf {\bibinfo {volume} {18}},\ \bibinfo {pages} {823} (\bibinfo {year}
  {1992})}\BibitemShut {NoStop}%
\bibitem [{\citenamefont {{Barry R. Holstein}}\ and\ \citenamefont
  {Treiman}(1977)}]{holstein:77}%
  \BibitemOpen
  \bibfield  {author} {\bibinfo {author} {\bibnamefont {{Barry R. Holstein}}}\
  and\ \bibinfo {author} {\bibfnamefont {S.~B.}\ \bibnamefont {Treiman}},\
  }\href {\doibase 10.1103/PhysRevD.16.2369} {\bibfield  {journal} {\bibinfo
  {journal} {Phys. Rev. D}\ }\textbf {\bibinfo {volume} {16}},\ \bibinfo
  {pages} {2369} (\bibinfo {year} {1977})}\BibitemShut {NoStop}%
\bibitem [{\citenamefont {{Barry R. Holstein}}(1977)}]{holstein:fierz}%
  \BibitemOpen
  \bibfield  {author} {\bibinfo {author} {\bibnamefont {{Barry R. Holstein}}},\
  }\href {\doibase 10.1103/PhysRevC.16.753} {\bibfield  {journal} {\bibinfo
  {journal} {Phys. Rev. C}\ }\textbf {\bibinfo {volume} {16}},\ \bibinfo
  {pages} {753} (\bibinfo {year} {1977})}\BibitemShut {NoStop}%
\bibitem [{\citenamefont {Hallin}\ \emph {et~al.}(1984)\citenamefont {Hallin},
  \citenamefont {Calaprice}, \citenamefont {MacArthur}, \citenamefont
  {Piilonen}, \citenamefont {Schneider},\ and\ \citenamefont
  {Schreiber}}]{hallin:84}%
  \BibitemOpen
  \bibfield  {author} {\bibinfo {author} {\bibfnamefont {A.~L.}\ \bibnamefont
  {Hallin}}, \bibinfo {author} {\bibfnamefont {F.~P.}\ \bibnamefont
  {Calaprice}}, \bibinfo {author} {\bibfnamefont {D.~W.}\ \bibnamefont
  {MacArthur}}, \bibinfo {author} {\bibfnamefont {L.~E.}\ \bibnamefont
  {Piilonen}}, \bibinfo {author} {\bibfnamefont {M.~B.}\ \bibnamefont
  {Schneider}}, \ and\ \bibinfo {author} {\bibfnamefont {D.~F.}\ \bibnamefont
  {Schreiber}},\ }\href {\doibase 10.1103/PhysRevLett.52.337} {\bibfield
  {journal} {\bibinfo  {journal} {Phys. Rev. Lett.}\ }\textbf {\bibinfo
  {volume} {52}},\ \bibinfo {pages} {337} (\bibinfo {year} {1984})}\BibitemShut
  {NoStop}%
\bibitem [{\citenamefont {Schneider}\ \emph {et~al.}(1983)\citenamefont
  {Schneider}, \citenamefont {Calaprice}, \citenamefont {Hallin}, \citenamefont
  {MacArthur},\ and\ \citenamefont {Schreiber}}]{schneider:83}%
  \BibitemOpen
  \bibfield  {author} {\bibinfo {author} {\bibfnamefont {M.~B.}\ \bibnamefont
  {Schneider}}, \bibinfo {author} {\bibfnamefont {F.~P.}\ \bibnamefont
  {Calaprice}}, \bibinfo {author} {\bibfnamefont {A.~L.}\ \bibnamefont
  {Hallin}}, \bibinfo {author} {\bibfnamefont {D.~W.}\ \bibnamefont
  {MacArthur}}, \ and\ \bibinfo {author} {\bibfnamefont {D.~F.}\ \bibnamefont
  {Schreiber}},\ }\href {\doibase 10.1103/PhysRevLett.51.1239} {\bibfield
  {journal} {\bibinfo  {journal} {Phys. Rev. Lett.}\ }\textbf {\bibinfo
  {volume} {51}},\ \bibinfo {pages} {1239} (\bibinfo {year}
  {1983})}\BibitemShut {NoStop}%
\bibitem [{\citenamefont {Naviliat-Cuncic}\ and\ \citenamefont
  {Severijns}(2009)}]{oscar:09}%
  \BibitemOpen
  \bibfield  {author} {\bibinfo {author} {\bibfnamefont {O.}~\bibnamefont
  {Naviliat-Cuncic}}\ and\ \bibinfo {author} {\bibfnamefont {N.}~\bibnamefont
  {Severijns}},\ }\href {\doibase 10.1103/PhysRevLett.102.142302} {\bibfield
  {journal} {\bibinfo  {journal} {Phys. Rev. Lett.}\ }\textbf {\bibinfo
  {volume} {102}},\ \bibinfo {pages} {142302} (\bibinfo {year}
  {2009})}\BibitemShut {NoStop}%
\bibitem [{\citenamefont {Haxton}\ \emph {et~al.}(1980)\citenamefont {Haxton},
  \citenamefont {Gibson},\ and\ \citenamefont {Henley}}]{haxton:80}%
  \BibitemOpen
  \bibfield  {author} {\bibinfo {author} {\bibfnamefont {W.~C.}\ \bibnamefont
  {Haxton}}, \bibinfo {author} {\bibfnamefont {B.~F.}\ \bibnamefont {Gibson}},
  \ and\ \bibinfo {author} {\bibfnamefont {E.~M.}\ \bibnamefont {Henley}},\
  }\href {\doibase 10.1103/PhysRevLett.45.1677} {\bibfield  {journal} {\bibinfo
   {journal} {Phys. Rev. Lett.}\ }\textbf {\bibinfo {volume} {45}},\ \bibinfo
  {pages} {1677} (\bibinfo {year} {1980})}\BibitemShut {NoStop}%
\bibitem [{\citenamefont {Brown}\ \emph {et~al.}(1980)\citenamefont {Brown},
  \citenamefont {Richter},\ and\ \citenamefont {Godwin}}]{brown:80}%
  \BibitemOpen
  \bibfield  {author} {\bibinfo {author} {\bibfnamefont {B.~A.}\ \bibnamefont
  {Brown}}, \bibinfo {author} {\bibfnamefont {W.~A.}\ \bibnamefont {Richter}},
  \ and\ \bibinfo {author} {\bibfnamefont {N.~S.}\ \bibnamefont {Godwin}},\
  }\href {\doibase 10.1103/PhysRevLett.45.1681} {\bibfield  {journal} {\bibinfo
   {journal} {Phys. Rev. Lett.}\ }\textbf {\bibinfo {volume} {45}},\ \bibinfo
  {pages} {1681} (\bibinfo {year} {1980})}\BibitemShut {NoStop}%
\bibitem [{\citenamefont {Adelberger}\ \emph {et~al.}(1983)\citenamefont
  {Adelberger}, \citenamefont {Hindi}, \citenamefont {Hoyle}, \citenamefont
  {Swanson}, \citenamefont {Von~Lintig},\ and\ \citenamefont
  {Haxton}}]{ega:83}%
  \BibitemOpen
  \bibfield  {author} {\bibinfo {author} {\bibfnamefont {E.~G.}\ \bibnamefont
  {Adelberger}}, \bibinfo {author} {\bibfnamefont {M.~M.}\ \bibnamefont
  {Hindi}}, \bibinfo {author} {\bibfnamefont {C.~D.}\ \bibnamefont {Hoyle}},
  \bibinfo {author} {\bibfnamefont {H.~E.}\ \bibnamefont {Swanson}}, \bibinfo
  {author} {\bibfnamefont {R.~D.}\ \bibnamefont {Von~Lintig}}, \ and\ \bibinfo
  {author} {\bibfnamefont {W.~C.}\ \bibnamefont {Haxton}},\ }\href {\doibase
  10.1103/PhysRevC.27.2833} {\bibfield  {journal} {\bibinfo  {journal} {Phys.
  Rev. C}\ }\textbf {\bibinfo {volume} {27}},\ \bibinfo {pages} {2833}
  (\bibinfo {year} {1983})}\BibitemShut {NoStop}%
\bibitem [{\citenamefont {Adelberger}\ and\ \citenamefont
  {Haxton}(1985)}]{Adelberger:85}%
  \BibitemOpen
  \bibfield  {author} {\bibinfo {author} {\bibfnamefont {E.~G.}\ \bibnamefont
  {Adelberger}}\ and\ \bibinfo {author} {\bibfnamefont {W.~C.}\ \bibnamefont
  {Haxton}},\ }\href {\doibase 10.1146/annurev.ns.35.120185.002441} {\bibfield
  {journal} {\bibinfo  {journal} {Ann. Rev. Nucl. Part. Sci.}\ }\textbf
  {\bibinfo {volume} {35}},\ \bibinfo {pages} {501} (\bibinfo {year}
  {1985})}\BibitemShut {NoStop}%
\bibitem [{\citenamefont {Garnsworthy}\ and\ \citenamefont
  {Garrett}(2014)}]{Garnsworthy:14}%
  \BibitemOpen
  \bibfield  {author} {\bibinfo {author} {\bibfnamefont {A.~B.}\ \bibnamefont
  {Garnsworthy}}\ and\ \bibinfo {author} {\bibfnamefont {P.~E.}\ \bibnamefont
  {Garrett}},\ }\href {\doibase 10.1007/s10751-013-0888-4} {\bibfield
  {journal} {\bibinfo  {journal} {Hyperfine Interactions}\ }\textbf {\bibinfo
  {volume} {225}},\ \bibinfo {pages} {121} (\bibinfo {year}
  {2014})}\BibitemShut {NoStop}%
\bibitem [{\citenamefont {Garrett}\ \emph {et~al.}(2015)\citenamefont {Garrett}
  \emph {et~al.}}]{Garrett:15}%
  \BibitemOpen
  \bibfield  {author} {\bibinfo {author} {\bibfnamefont {P.~E.}\ \bibnamefont
  {Garrett}} \emph {et~al.},\ }\href
  {http://stacks.iop.org/1742-6596/639/i=1/a=012006} {\bibfield  {journal}
  {\bibinfo  {journal} {Journal of Physics: Conference Series}\ }\textbf
  {\bibinfo {volume} {639}},\ \bibinfo {pages} {012006} (\bibinfo {year}
  {2015})}\BibitemShut {NoStop}%
\bibitem [{\citenamefont {Triambak}\ \emph {et~al.}(2012)\citenamefont
  {Triambak} \emph {et~al.}}]{Triambak:12}%
  \BibitemOpen
  \bibfield  {author} {\bibinfo {author} {\bibfnamefont {S.}~\bibnamefont
  {Triambak}} \emph {et~al.},\ }\href {\doibase 10.1103/PhysRevLett.109.042301}
  {\bibfield  {journal} {\bibinfo  {journal} {Phys. Rev. Lett.}\ }\textbf
  {\bibinfo {volume} {109}},\ \bibinfo {pages} {042301} (\bibinfo {year}
  {2012})}\BibitemShut {NoStop}%
\bibitem [{\citenamefont {Finlay}\ \emph {et~al.}(2008)\citenamefont {Finlay}
  \emph {et~al.}}]{finlay}%
  \BibitemOpen
  \bibfield  {author} {\bibinfo {author} {\bibfnamefont {P.}~\bibnamefont
  {Finlay}} \emph {et~al.},\ }\href {\doibase 10.1103/PhysRevC.78.025502}
  {\bibfield  {journal} {\bibinfo  {journal} {Phys. Rev. C}\ }\textbf {\bibinfo
  {volume} {78}},\ \bibinfo {pages} {025502} (\bibinfo {year}
  {2008})}\BibitemShut {NoStop}%
\bibitem [{\citenamefont {Leach}\ \emph {et~al.}(2008)\citenamefont {Leach}
  \emph {et~al.}}]{leach}%
  \BibitemOpen
  \bibfield  {author} {\bibinfo {author} {\bibfnamefont {K.~G.}\ \bibnamefont
  {Leach}} \emph {et~al.},\ }\href {\doibase 10.1103/PhysRevLett.100.192504}
  {\bibfield  {journal} {\bibinfo  {journal} {Phys. Rev. Lett.}\ }\textbf
  {\bibinfo {volume} {100}},\ \bibinfo {pages} {192504} (\bibinfo {year}
  {2008})}\BibitemShut {NoStop}%
\bibitem [{\citenamefont {Dunlop}\ \emph {et~al.}(2013)\citenamefont {Dunlop}
  \emph {et~al.}}]{dunlop}%
  \BibitemOpen
  \bibfield  {author} {\bibinfo {author} {\bibfnamefont {R.}~\bibnamefont
  {Dunlop}} \emph {et~al.},\ }\href {\doibase 10.1103/PhysRevC.88.045501}
  {\bibfield  {journal} {\bibinfo  {journal} {Phys. Rev. C}\ }\textbf {\bibinfo
  {volume} {88}},\ \bibinfo {pages} {045501} (\bibinfo {year}
  {2013})}\BibitemShut {NoStop}%
\bibitem [{\citenamefont {Sempau}\ \emph {et~al.}(1997)\citenamefont {Sempau},
  \citenamefont {Acosta}, \citenamefont {Baro}, \citenamefont
  {Fern\'andez-Varea},\ and\ \citenamefont {Salvat}}]{penelope}%
  \BibitemOpen
  \bibfield  {author} {\bibinfo {author} {\bibfnamefont {J.}~\bibnamefont
  {Sempau}}, \bibinfo {author} {\bibfnamefont {E.}~\bibnamefont {Acosta}},
  \bibinfo {author} {\bibfnamefont {J.}~\bibnamefont {Baro}}, \bibinfo {author}
  {\bibfnamefont {J.}~\bibnamefont {Fern\'andez-Varea}}, \ and\ \bibinfo
  {author} {\bibfnamefont {F.}~\bibnamefont {Salvat}},\ }\href {\doibase
  https://doi.org/10.1016/S0168-583X(97)00414-X} {\bibfield  {journal}
  {\bibinfo  {journal} {Nucl. Inst. Meth. Phys. Res. B}\ }\textbf {\bibinfo
  {volume} {132}},\ \bibinfo {pages} {377 } (\bibinfo {year}
  {1997})}\BibitemShut {NoStop}%
\bibitem [{sri()}]{srim}%
  \BibitemOpen
  \href@noop {} {}\bibinfo {howpublished}
  {\url{http://www.srim.org/}}\BibitemShut {NoStop}%
\bibitem [{\citenamefont {{Ziegler}}\ \emph {et~al.}(2010)\citenamefont
  {{Ziegler}}, \citenamefont {{Ziegler}},\ and\ \citenamefont
  {{Biersack}}}]{srim2}%
  \BibitemOpen
  \bibfield  {author} {\bibinfo {author} {\bibfnamefont {J.~F.}\ \bibnamefont
  {{Ziegler}}}, \bibinfo {author} {\bibfnamefont {M.~D.}\ \bibnamefont
  {{Ziegler}}}, \ and\ \bibinfo {author} {\bibfnamefont {J.~P.}\ \bibnamefont
  {{Biersack}}},\ }\href {\doibase 10.1016/j.nimb.2010.02.091} {\bibfield
  {journal} {\bibinfo  {journal} {Nucl. Instr. Meth. Phys. Res. B}\ }\textbf
  {\bibinfo {volume} {268}},\ \bibinfo {pages} {1818} (\bibinfo {year}
  {2010})}\BibitemShut {NoStop}%
\bibitem [{\citenamefont {Iacob}\ \emph {et~al.}(2006)\citenamefont {Iacob},
  \citenamefont {Hardy}, \citenamefont {Gagliardi}, \citenamefont {Goodwin},
  \citenamefont {Nica}, \citenamefont {Park}, \citenamefont {Tabacaru},
  \citenamefont {Trache}, \citenamefont {Tribble}, \citenamefont {Zhai},\ and\
  \citenamefont {Towner}}]{iacob:06}%
  \BibitemOpen
  \bibfield  {author} {\bibinfo {author} {\bibfnamefont {V.~E.}\ \bibnamefont
  {Iacob}}, \bibinfo {author} {\bibfnamefont {J.~C.}\ \bibnamefont {Hardy}},
  \bibinfo {author} {\bibfnamefont {C.~A.}\ \bibnamefont {Gagliardi}}, \bibinfo
  {author} {\bibfnamefont {J.}~\bibnamefont {Goodwin}}, \bibinfo {author}
  {\bibfnamefont {N.}~\bibnamefont {Nica}}, \bibinfo {author} {\bibfnamefont
  {H.~I.}\ \bibnamefont {Park}}, \bibinfo {author} {\bibfnamefont
  {G.}~\bibnamefont {Tabacaru}}, \bibinfo {author} {\bibfnamefont
  {L.}~\bibnamefont {Trache}}, \bibinfo {author} {\bibfnamefont {R.~E.}\
  \bibnamefont {Tribble}}, \bibinfo {author} {\bibfnamefont {Y.}~\bibnamefont
  {Zhai}}, \ and\ \bibinfo {author} {\bibfnamefont {I.~S.}\ \bibnamefont
  {Towner}},\ }\href {\doibase 10.1103/PhysRevC.74.015501} {\bibfield
  {journal} {\bibinfo  {journal} {Phys. Rev. C}\ }\textbf {\bibinfo {volume}
  {74}},\ \bibinfo {pages} {015501} (\bibinfo {year} {2006})}\BibitemShut
  {NoStop}%
\bibitem [{\citenamefont {Baker}\ and\ \citenamefont
  {Cousins}(1984)}]{Baker:84}%
  \BibitemOpen
  \bibfield  {author} {\bibinfo {author} {\bibfnamefont {S.}~\bibnamefont
  {Baker}}\ and\ \bibinfo {author} {\bibfnamefont {R.~D.}\ \bibnamefont
  {Cousins}},\ }\href {\doibase https://doi.org/10.1016/0167-5087(84)90016-4}
  {\bibfield  {journal} {\bibinfo  {journal} {Nucl. Instr. Meth. Phys. Res.}\
  }\textbf {\bibinfo {volume} {221}},\ \bibinfo {pages} {437 } (\bibinfo {year}
  {1984})}\BibitemShut {NoStop}%
\bibitem [{\citenamefont {Uji\ifmmode~\acute{c}\else \'{c}\fi{}}\ \emph
  {et~al.}(2013)\citenamefont {Uji\ifmmode~\acute{c}\else \'{c}\fi{}} \emph
  {et~al.}}]{Ujic:13}%
  \BibitemOpen
  \bibfield  {author} {\bibinfo {author} {\bibfnamefont {P.}~\bibnamefont
  {Uji\ifmmode~\acute{c}\else \'{c}\fi{}}} \emph {et~al.},\ }\href {\doibase
  10.1103/PhysRevLett.110.032501} {\bibfield  {journal} {\bibinfo  {journal}
  {Phys. Rev. Lett.}\ }\textbf {\bibinfo {volume} {110}},\ \bibinfo {pages}
  {032501} (\bibinfo {year} {2013})}\BibitemShut {NoStop}%
\bibitem [{\citenamefont {Fontbonne}\ \emph {et~al.}(2017)\citenamefont
  {Fontbonne} \emph {et~al.}}]{Fontbonne:17}%
  \BibitemOpen
  \bibfield  {author} {\bibinfo {author} {\bibfnamefont {C.}~\bibnamefont
  {Fontbonne}} \emph {et~al.},\ }\href {\doibase 10.1103/PhysRevC.96.065501}
  {\bibfield  {journal} {\bibinfo  {journal} {Phys. Rev. C}\ }\textbf {\bibinfo
  {volume} {96}},\ \bibinfo {pages} {065501} (\bibinfo {year}
  {2017})}\BibitemShut {NoStop}%
\bibitem [{\citenamefont {Broussard}\ \emph {et~al.}(2014)\citenamefont
  {Broussard} \emph {et~al.}}]{Broussard:14}%
  \BibitemOpen
  \bibfield  {author} {\bibinfo {author} {\bibfnamefont {L.~J.}\ \bibnamefont
  {Broussard}} \emph {et~al.},\ }\href {\doibase
  10.1103/PhysRevLett.112.212301} {\bibfield  {journal} {\bibinfo  {journal}
  {Phys. Rev. Lett.}\ }\textbf {\bibinfo {volume} {112}},\ \bibinfo {pages}
  {212301} (\bibinfo {year} {2014})}\BibitemShut {NoStop}%
\bibitem [{\citenamefont {{G.F. Grinyer}}\ \emph {et~al.}(2007)\citenamefont
  {{G.F. Grinyer}} \emph {et~al.}}]{geoff:nim}%
  \BibitemOpen
  \bibfield  {author} {\bibinfo {author} {\bibnamefont {{G.F. Grinyer}}} \emph
  {et~al.},\ }\href {\doibase https://doi.org/10.1016/j.nima.2007.05.323}
  {\bibfield  {journal} {\bibinfo  {journal} {Nucl. Instr. Meth. Phys. Res. A}\
  }\textbf {\bibinfo {volume} {579}},\ \bibinfo {pages} {1005 } (\bibinfo
  {year} {2007})}\BibitemShut {NoStop}%
\bibitem [{\citenamefont {Finlay}(2007)}]{finlay_thesis}%
  \BibitemOpen
  \bibfield  {author} {\bibinfo {author} {\bibfnamefont {P.}~\bibnamefont
  {Finlay}},\ }\href@noop {} {\bibinfo {type} {{M.Sc.} thesis}},\ \bibinfo
  {school} {University of Guelph} (\bibinfo {year} {2007})\BibitemShut
  {NoStop}%
\bibitem [{fin()}]{finlay2}%
  \BibitemOpen
  \href@noop {} {}\bibinfo {howpublished}
  {\url{https://www.physics.uoguelph.ca/Nucweb/theses/paulfinlay_62Ga_mscthesis.pdf}}\BibitemShut
  {NoStop}%
\bibitem [{\citenamefont {Freedman}\ \emph {et~al.}(1975)\citenamefont
  {Freedman}, \citenamefont {Del~Vecchio},\ and\ \citenamefont
  {Callias}}]{Freedman:75}%
  \BibitemOpen
  \bibfield  {author} {\bibinfo {author} {\bibfnamefont {S.~J.}\ \bibnamefont
  {Freedman}}, \bibinfo {author} {\bibfnamefont {R.~M.}\ \bibnamefont
  {Del~Vecchio}}, \ and\ \bibinfo {author} {\bibfnamefont {C.}~\bibnamefont
  {Callias}},\ }\href {\doibase 10.1103/PhysRevC.12.315} {\bibfield  {journal}
  {\bibinfo  {journal} {Phys. Rev. C}\ }\textbf {\bibinfo {volume} {12}},\
  \bibinfo {pages} {315} (\bibinfo {year} {1975})}\BibitemShut {NoStop}%
\bibitem [{\citenamefont {Severijns}\ \emph {et~al.}(2008)\citenamefont
  {Severijns}, \citenamefont {Tandecki}, \citenamefont {Phalet},\ and\
  \citenamefont {Towner}}]{nathal_prc}%
  \BibitemOpen
  \bibfield  {author} {\bibinfo {author} {\bibfnamefont {N.}~\bibnamefont
  {Severijns}}, \bibinfo {author} {\bibfnamefont {M.}~\bibnamefont {Tandecki}},
  \bibinfo {author} {\bibfnamefont {T.}~\bibnamefont {Phalet}}, \ and\ \bibinfo
  {author} {\bibfnamefont {I.~S.}\ \bibnamefont {Towner}},\ }\href {\doibase
  10.1103/PhysRevC.78.055501} {\bibfield  {journal} {\bibinfo  {journal} {Phys.
  Rev. C}\ }\textbf {\bibinfo {volume} {78}},\ \bibinfo {pages} {055501}
  (\bibinfo {year} {2008})}\BibitemShut {NoStop}%
\bibitem [{\citenamefont {Alburger}(1976)}]{Alb:76}%
  \BibitemOpen
  \bibfield  {author} {\bibinfo {author} {\bibfnamefont {D.~E.}\ \bibnamefont
  {Alburger}},\ }\href {\doibase 10.1103/PhysRevC.13.2593} {\bibfield
  {journal} {\bibinfo  {journal} {Phys. Rev. C}\ }\textbf {\bibinfo {volume}
  {13}},\ \bibinfo {pages} {2593} (\bibinfo {year} {1976})}\BibitemShut
  {NoStop}%
\bibitem [{\citenamefont {Adelberger}\ \emph {et~al.}(1981)\citenamefont
  {Adelberger}, \citenamefont {Hindi}, \citenamefont {Hoyle}, \citenamefont
  {Swanson},\ and\ \citenamefont {Von~Lintig}}]{Adel:81}%
  \BibitemOpen
  \bibfield  {author} {\bibinfo {author} {\bibfnamefont {E.~G.}\ \bibnamefont
  {Adelberger}}, \bibinfo {author} {\bibfnamefont {M.~M.}\ \bibnamefont
  {Hindi}}, \bibinfo {author} {\bibfnamefont {C.~D.}\ \bibnamefont {Hoyle}},
  \bibinfo {author} {\bibfnamefont {H.~E.}\ \bibnamefont {Swanson}}, \ and\
  \bibinfo {author} {\bibfnamefont {R.~D.}\ \bibnamefont {Von~Lintig}},\ }\href
  {\doibase 10.1103/PhysRevC.24.313} {\bibfield  {journal} {\bibinfo  {journal}
  {Phys. Rev. C}\ }\textbf {\bibinfo {volume} {24}},\ \bibinfo {pages} {313}
  (\bibinfo {year} {1981})}\BibitemShut {NoStop}%
\bibitem [{\citenamefont {Saettler}\ \emph {et~al.}(1993)\citenamefont
  {Saettler}, \citenamefont {Calaprice}, \citenamefont {Hallin},\ and\
  \citenamefont {Lowry}}]{Saett:93}%
  \BibitemOpen
  \bibfield  {author} {\bibinfo {author} {\bibfnamefont {E.~R.~J.}\
  \bibnamefont {Saettler}}, \bibinfo {author} {\bibfnamefont {F.~P.}\
  \bibnamefont {Calaprice}}, \bibinfo {author} {\bibfnamefont {A.~L.}\
  \bibnamefont {Hallin}}, \ and\ \bibinfo {author} {\bibfnamefont {M.~M.}\
  \bibnamefont {Lowry}},\ }\href {\doibase 10.1103/PhysRevC.48.3069} {\bibfield
   {journal} {\bibinfo  {journal} {Phys. Rev. C}\ }\textbf {\bibinfo {volume}
  {48}},\ \bibinfo {pages} {3069} (\bibinfo {year} {1993})}\BibitemShut
  {NoStop}%
\bibitem [{\citenamefont {Firestone}\ \emph {et~al.}(1966)\citenamefont
  {Firestone} \emph {et~al.}}]{toi}%
  \BibitemOpen
  \bibinfo {editor} {\bibfnamefont {R.~B.}\ \bibnamefont {Firestone}} \emph
  {et~al.},\ eds.,\ in\ \href@noop {} {\emph {\bibinfo {booktitle} {Table of
  isotopes}}}\ (\bibinfo  {publisher} {Wiley},\ \bibinfo {address} {New York},\
  \bibinfo {year} {1966})\ \bibinfo {edition} {8th}\ ed.\BibitemShut {Stop}%
\bibitem [{\citenamefont {Bambynek}\ \emph {et~al.}(1977)\citenamefont
  {Bambynek} \emph {et~al.}}]{bamb}%
  \BibitemOpen
  \bibfield  {author} {\bibinfo {author} {\bibfnamefont {W.}~\bibnamefont
  {Bambynek}} \emph {et~al.},\ }\href {\doibase 10.1103/RevModPhys.49.77}
  {\bibfield  {journal} {\bibinfo  {journal} {Rev. Mod. Phys.}\ }\textbf
  {\bibinfo {volume} {49}},\ \bibinfo {pages} {77} (\bibinfo {year}
  {1977})}\BibitemShut {NoStop}%
\bibitem [{amd()}]{amdcweb}%
  \BibitemOpen
  \href@noop {} {}\bibinfo {howpublished}
  {\url{https://www-nds.iaea.org/amdc/}}\BibitemShut {NoStop}%
\bibitem [{\citenamefont {Wang}\ \emph {et~al.}(2017)\citenamefont {Wang},
  \citenamefont {Audi}, \citenamefont {Kondev}, \citenamefont {Huang},
  \citenamefont {Naimi},\ and\ \citenamefont {Xu}}]{amdc:16}%
  \BibitemOpen
  \bibfield  {author} {\bibinfo {author} {\bibfnamefont {M.}~\bibnamefont
  {Wang}}, \bibinfo {author} {\bibfnamefont {G.}~\bibnamefont {Audi}}, \bibinfo
  {author} {\bibfnamefont {F.}~\bibnamefont {Kondev}}, \bibinfo {author}
  {\bibfnamefont {W.}~\bibnamefont {Huang}}, \bibinfo {author} {\bibfnamefont
  {S.}~\bibnamefont {Naimi}}, \ and\ \bibinfo {author} {\bibfnamefont
  {X.}~\bibnamefont {Xu}},\ }\href
  {http://stacks.iop.org/1674-1137/41/i=3/a=030003} {\bibfield  {journal}
  {\bibinfo  {journal} {Chinese Physics C}\ }\textbf {\bibinfo {volume} {41}},\
  \bibinfo {pages} {030003} (\bibinfo {year} {2017})}\BibitemShut {NoStop}%
\bibitem [{\citenamefont {Hardy}\ and\ \citenamefont {Towner}(2015)}]{TH:15}%
  \BibitemOpen
  \bibfield  {author} {\bibinfo {author} {\bibfnamefont {J.~C.}\ \bibnamefont
  {Hardy}}\ and\ \bibinfo {author} {\bibfnamefont {I.~S.}\ \bibnamefont
  {Towner}},\ }\href {\doibase 10.1103/PhysRevC.91.025501} {\bibfield
  {journal} {\bibinfo  {journal} {Phys. Rev. C}\ }\textbf {\bibinfo {volume}
  {91}},\ \bibinfo {pages} {025501} (\bibinfo {year} {2015})}\BibitemShut
  {NoStop}%
\bibitem [{\citenamefont {Triambak}\ \emph {et~al.}(2017)\citenamefont
  {Triambak} \emph {et~al.}}]{Triambak:17}%
  \BibitemOpen
  \bibfield  {author} {\bibinfo {author} {\bibfnamefont {S.}~\bibnamefont
  {Triambak}} \emph {et~al.},\ }\href {\doibase 10.1103/PhysRevC.95.035501}
  {\bibfield  {journal} {\bibinfo  {journal} {Phys. Rev. C}\ }\textbf {\bibinfo
  {volume} {95}},\ \bibinfo {pages} {035501} (\bibinfo {year}
  {2017})}\BibitemShut {NoStop}%
\bibitem [{\citenamefont {Grenacs}(1985)}]{Grenacs}%
  \BibitemOpen
  \bibfield  {author} {\bibinfo {author} {\bibfnamefont {L.}~\bibnamefont
  {Grenacs}},\ }\href {\doibase 10.1146/annurev.ns.35.120185.002323} {\bibfield
   {journal} {\bibinfo  {journal} {Ann. Rev. Nucl. Part. Sci.}\ }\textbf
  {\bibinfo {volume} {35}},\ \bibinfo {pages} {455} (\bibinfo {year}
  {1985})}\BibitemShut {NoStop}%
\bibitem [{ian()}]{ian_pvt}%
  \BibitemOpen
  \href@noop {} {}\bibinfo {howpublished} {I. S. Towner, private
  communication.}\BibitemShut {Stop}%
\bibitem [{\citenamefont {{Barry R.
  Holstein}}(1974{\natexlab{a}})}]{Holstein:rmp}%
  \BibitemOpen
  \bibfield  {author} {\bibinfo {author} {\bibnamefont {{Barry R. Holstein}}},\
  }\href {\doibase 10.1103/RevModPhys.46.789} {\bibfield  {journal} {\bibinfo
  {journal} {Rev. Mod. Phys.}\ }\textbf {\bibinfo {volume} {46}},\ \bibinfo
  {pages} {789} (\bibinfo {year} {1974}{\natexlab{a}})}\BibitemShut {NoStop}%
\bibitem [{\citenamefont {{Vincenzo Cirigliano}}\ \emph
  {et~al.}(2013)\citenamefont {{Vincenzo Cirigliano}}, \citenamefont {{Susan
  Gardner}},\ and\ \citenamefont {{Barry R. Holstein}}}]{crigliano}%
  \BibitemOpen
  \bibfield  {author} {\bibinfo {author} {\bibnamefont {{Vincenzo
  Cirigliano}}}, \bibinfo {author} {\bibnamefont {{Susan Gardner}}}, \ and\
  \bibinfo {author} {\bibnamefont {{Barry R. Holstein}}},\ }\href {\doibase
  https://doi.org/10.1016/j.ppnp.2013.03.005} {\bibfield  {journal} {\bibinfo
  {journal} {Progress in Particle and Nuclear Physics}\ }\textbf {\bibinfo
  {volume} {71}},\ \bibinfo {pages} {93 } (\bibinfo {year} {2013})}\BibitemShut
  {NoStop}%
\bibitem [{\citenamefont {Weinberg}(1958)}]{Weinberg}%
  \BibitemOpen
  \bibfield  {author} {\bibinfo {author} {\bibfnamefont {S.}~\bibnamefont
  {Weinberg}},\ }\href {\doibase 10.1103/PhysRev.112.1375} {\bibfield
  {journal} {\bibinfo  {journal} {Phys. Rev.}\ }\textbf {\bibinfo {volume}
  {112}},\ \bibinfo {pages} {1375} (\bibinfo {year} {1958})}\BibitemShut
  {NoStop}%
\bibitem [{\citenamefont {{Barry R. Holstein}}\ and\ \citenamefont
  {Treiman}(1971)}]{Holstein:71}%
  \BibitemOpen
  \bibfield  {author} {\bibinfo {author} {\bibnamefont {{Barry R. Holstein}}}\
  and\ \bibinfo {author} {\bibfnamefont {S.~B.}\ \bibnamefont {Treiman}},\
  }\href {\doibase 10.1103/PhysRevC.3.1921} {\bibfield  {journal} {\bibinfo
  {journal} {Phys. Rev. C}\ }\textbf {\bibinfo {volume} {3}},\ \bibinfo {pages}
  {1921} (\bibinfo {year} {1971})}\BibitemShut {NoStop}%
\bibitem [{\citenamefont {{Barry R.
  Holstein}}(1974{\natexlab{b}})}]{holstein:coul}%
  \BibitemOpen
  \bibfield  {author} {\bibinfo {author} {\bibnamefont {{Barry R. Holstein}}},\
  }\href {\doibase 10.1103/PhysRevC.9.1742} {\bibfield  {journal} {\bibinfo
  {journal} {Phys. Rev. C}\ }\textbf {\bibinfo {volume} {9}},\ \bibinfo {pages}
  {1742} (\bibinfo {year} {1974}{\natexlab{b}})}\BibitemShut {NoStop}%
\bibitem [{\citenamefont {Feynman}\ and\ \citenamefont
  {Gell-Mann}(1958)}]{cvc1}%
  \BibitemOpen
  \bibfield  {author} {\bibinfo {author} {\bibfnamefont {R.~P.}\ \bibnamefont
  {Feynman}}\ and\ \bibinfo {author} {\bibfnamefont {M.}~\bibnamefont
  {Gell-Mann}},\ }\href {\doibase 10.1103/PhysRev.109.193} {\bibfield
  {journal} {\bibinfo  {journal} {Phys. Rev.}\ }\textbf {\bibinfo {volume}
  {109}},\ \bibinfo {pages} {193} (\bibinfo {year} {1958})}\BibitemShut
  {NoStop}%
\bibitem [{\citenamefont {Commins}\ and\ \citenamefont
  {Dobson}(1963)}]{Commins:63}%
  \BibitemOpen
  \bibfield  {author} {\bibinfo {author} {\bibfnamefont {E.~D.}\ \bibnamefont
  {Commins}}\ and\ \bibinfo {author} {\bibfnamefont {D.~A.}\ \bibnamefont
  {Dobson}},\ }\href {\doibase 10.1103/PhysRevLett.10.347} {\bibfield
  {journal} {\bibinfo  {journal} {Phys. Rev. Lett.}\ }\textbf {\bibinfo
  {volume} {10}},\ \bibinfo {pages} {347} (\bibinfo {year} {1963})}\BibitemShut
  {NoStop}%
\bibitem [{\citenamefont {Calaprice}\ \emph {et~al.}(1967)\citenamefont
  {Calaprice}, \citenamefont {Commins}, \citenamefont {Gibbs}, \citenamefont
  {Wick},\ and\ \citenamefont {Dobson}}]{Calaprice:67}%
  \BibitemOpen
  \bibfield  {author} {\bibinfo {author} {\bibfnamefont {F.~P.}\ \bibnamefont
  {Calaprice}}, \bibinfo {author} {\bibfnamefont {E.~D.}\ \bibnamefont
  {Commins}}, \bibinfo {author} {\bibfnamefont {H.~M.}\ \bibnamefont {Gibbs}},
  \bibinfo {author} {\bibfnamefont {G.~L.}\ \bibnamefont {Wick}}, \ and\
  \bibinfo {author} {\bibfnamefont {D.~A.}\ \bibnamefont {Dobson}},\ }\href
  {\doibase 10.1103/PhysRevLett.18.918} {\bibfield  {journal} {\bibinfo
  {journal} {Phys. Rev. Lett.}\ }\textbf {\bibinfo {volume} {18}},\ \bibinfo
  {pages} {918} (\bibinfo {year} {1967})}\BibitemShut {NoStop}%
\bibitem [{\citenamefont {Calaprice}\ \emph {et~al.}(1969)\citenamefont
  {Calaprice}, \citenamefont {Commins}, \citenamefont {Gibbs}, \citenamefont
  {Wick},\ and\ \citenamefont {Dobson}}]{Calaprice:69}%
  \BibitemOpen
  \bibfield  {author} {\bibinfo {author} {\bibfnamefont {F.~P.}\ \bibnamefont
  {Calaprice}}, \bibinfo {author} {\bibfnamefont {E.~D.}\ \bibnamefont
  {Commins}}, \bibinfo {author} {\bibfnamefont {H.~M.}\ \bibnamefont {Gibbs}},
  \bibinfo {author} {\bibfnamefont {G.~L.}\ \bibnamefont {Wick}}, \ and\
  \bibinfo {author} {\bibfnamefont {D.~A.}\ \bibnamefont {Dobson}},\ }\href
  {\doibase 10.1103/PhysRev.184.1117} {\bibfield  {journal} {\bibinfo
  {journal} {Phys. Rev.}\ }\textbf {\bibinfo {volume} {184}},\ \bibinfo {pages}
  {1117} (\bibinfo {year} {1969})}\BibitemShut {NoStop}%
\bibitem [{\citenamefont {Schreiber}(1983)}]{Schreiber}%
  \BibitemOpen
  \bibfield  {author} {\bibinfo {author} {\bibfnamefont {D.~F.}\ \bibnamefont
  {Schreiber}},\ }\href@noop {} {\bibinfo {type} {{Ph.D.} thesis}},\ \bibinfo
  {school} {Princeton University} (\bibinfo {year} {1983})\BibitemShut
  {NoStop}%
\bibitem [{\citenamefont {Jones}(1996)}]{Jones}%
  \BibitemOpen
  \bibfield  {author} {\bibinfo {author} {\bibfnamefont {G.~L.}\ \bibnamefont
  {Jones}},\ }\href@noop {} {\bibinfo {type} {{Ph.D.} thesis}},\ \bibinfo
  {school} {Princeton University} (\bibinfo {year} {1996})\BibitemShut
  {NoStop}%
\bibitem [{\citenamefont {Tribble}\ \emph {et~al.}(1981)\citenamefont
  {Tribble}, \citenamefont {May},\ and\ \citenamefont {Tanner}}]{tribble1}%
  \BibitemOpen
  \bibfield  {author} {\bibinfo {author} {\bibfnamefont {R.~E.}\ \bibnamefont
  {Tribble}}, \bibinfo {author} {\bibfnamefont {D.~P.}\ \bibnamefont {May}}, \
  and\ \bibinfo {author} {\bibfnamefont {D.~M.}\ \bibnamefont {Tanner}},\
  }\href {\doibase 10.1103/PhysRevC.23.2245} {\bibfield  {journal} {\bibinfo
  {journal} {Phys. Rev. C}\ }\textbf {\bibinfo {volume} {23}},\ \bibinfo
  {pages} {2245} (\bibinfo {year} {1981})}\BibitemShut {NoStop}%
\bibitem [{\citenamefont {Tribble}\ and\ \citenamefont {May}(1978)}]{tribble2}%
  \BibitemOpen
  \bibfield  {author} {\bibinfo {author} {\bibfnamefont {R.~E.}\ \bibnamefont
  {Tribble}}\ and\ \bibinfo {author} {\bibfnamefont {D.~P.}\ \bibnamefont
  {May}},\ }\href {\doibase 10.1103/PhysRevC.18.2704} {\bibfield  {journal}
  {\bibinfo  {journal} {Phys. Rev. C}\ }\textbf {\bibinfo {volume} {18}},\
  \bibinfo {pages} {2704} (\bibinfo {year} {1978})}\BibitemShut {NoStop}%
\bibitem [{\citenamefont {Minamisono}\ \emph {et~al.}(2011)\citenamefont
  {Minamisono} \emph {et~al.}}]{kei1}%
  \BibitemOpen
  \bibfield  {author} {\bibinfo {author} {\bibfnamefont {K.}~\bibnamefont
  {Minamisono}} \emph {et~al.},\ }\href {\doibase 10.1103/PhysRevC.84.055501}
  {\bibfield  {journal} {\bibinfo  {journal} {Phys. Rev. C}\ }\textbf {\bibinfo
  {volume} {84}},\ \bibinfo {pages} {055501} (\bibinfo {year}
  {2011})}\BibitemShut {NoStop}%
\bibitem [{\citenamefont {Minamisono}\ \emph {et~al.}(2001)\citenamefont
  {Minamisono}, \citenamefont {Matsuta}, \citenamefont {Minamisono},
  \citenamefont {Yamaguchi}, \citenamefont {Sumikama}, \citenamefont
  {Nagatomo}, \citenamefont {Ogura}, \citenamefont {Iwakoshi}, \citenamefont
  {Fukuda}, \citenamefont {Mihara}, \citenamefont {Koshigiri},\ and\
  \citenamefont {Morita}}]{kei2}%
  \BibitemOpen
  \bibfield  {author} {\bibinfo {author} {\bibfnamefont {K.}~\bibnamefont
  {Minamisono}}, \bibinfo {author} {\bibfnamefont {K.}~\bibnamefont {Matsuta}},
  \bibinfo {author} {\bibfnamefont {T.}~\bibnamefont {Minamisono}}, \bibinfo
  {author} {\bibfnamefont {T.}~\bibnamefont {Yamaguchi}}, \bibinfo {author}
  {\bibfnamefont {T.}~\bibnamefont {Sumikama}}, \bibinfo {author}
  {\bibfnamefont {T.}~\bibnamefont {Nagatomo}}, \bibinfo {author}
  {\bibfnamefont {M.}~\bibnamefont {Ogura}}, \bibinfo {author} {\bibfnamefont
  {T.}~\bibnamefont {Iwakoshi}}, \bibinfo {author} {\bibfnamefont
  {M.}~\bibnamefont {Fukuda}}, \bibinfo {author} {\bibfnamefont
  {M.}~\bibnamefont {Mihara}}, \bibinfo {author} {\bibfnamefont
  {K.}~\bibnamefont {Koshigiri}}, \ and\ \bibinfo {author} {\bibfnamefont
  {M.}~\bibnamefont {Morita}},\ }\href {\doibase 10.1103/PhysRevC.65.015501}
  {\bibfield  {journal} {\bibinfo  {journal} {Phys. Rev. C}\ }\textbf {\bibinfo
  {volume} {65}},\ \bibinfo {pages} {015501} (\bibinfo {year}
  {2001})}\BibitemShut {NoStop}%
\bibitem [{\citenamefont {Pati}\ and\ \citenamefont {Salam}(1974)}]{pati:74}%
  \BibitemOpen
  \bibfield  {author} {\bibinfo {author} {\bibfnamefont {J.~C.}\ \bibnamefont
  {Pati}}\ and\ \bibinfo {author} {\bibfnamefont {A.}~\bibnamefont {Salam}},\
  }\href {\doibase 10.1103/PhysRevD.10.275} {\bibfield  {journal} {\bibinfo
  {journal} {Phys. Rev. D}\ }\textbf {\bibinfo {volume} {10}},\ \bibinfo
  {pages} {275} (\bibinfo {year} {1974})}\BibitemShut {NoStop}%
\bibitem [{\citenamefont {Senjanovi\'c}\ and\ \citenamefont
  {Mohapatra}(1975)}]{senjanovic:75}%
  \BibitemOpen
  \bibfield  {author} {\bibinfo {author} {\bibfnamefont {G.}~\bibnamefont
  {Senjanovi\'c}}\ and\ \bibinfo {author} {\bibfnamefont {R.~N.}\ \bibnamefont
  {Mohapatra}},\ }\href {\doibase 10.1103/PhysRevD.12.1502} {\bibfield
  {journal} {\bibinfo  {journal} {Phys. Rev. D}\ }\textbf {\bibinfo {volume}
  {12}},\ \bibinfo {pages} {1502} (\bibinfo {year} {1975})}\BibitemShut
  {NoStop}%
\bibitem [{\citenamefont {Shaban}\ and\ \citenamefont
  {Stirling}(1992)}]{Shaban}%
  \BibitemOpen
  \bibfield  {author} {\bibinfo {author} {\bibfnamefont {N.}~\bibnamefont
  {Shaban}}\ and\ \bibinfo {author} {\bibfnamefont {W.}~\bibnamefont
  {Stirling}},\ }\href {\doibase https://doi.org/10.1016/0370-2693(92)91046-C}
  {\bibfield  {journal} {\bibinfo  {journal} {Physics Letters B}\ }\textbf
  {\bibinfo {volume} {291}},\ \bibinfo {pages} {281 } (\bibinfo {year}
  {1992})}\BibitemShut {NoStop}%
\bibitem [{\citenamefont {Herczeg}(2001)}]{Herczeg}%
  \BibitemOpen
  \bibfield  {author} {\bibinfo {author} {\bibfnamefont {P.}~\bibnamefont
  {Herczeg}},\ }\href {\doibase https://doi.org/10.1016/S0146-6410(01)00149-1}
  {\bibfield  {journal} {\bibinfo  {journal} {Progress in Particle and Nuclear
  Physics}\ }\textbf {\bibinfo {volume} {46}},\ \bibinfo {pages} {413 }
  (\bibinfo {year} {2001})}\BibitemShut {NoStop}%
\bibitem [{\citenamefont {B\'eg}\ \emph {et~al.}(1977)\citenamefont {B\'eg},
  \citenamefont {Budny}, \citenamefont {Mohapatra},\ and\ \citenamefont
  {Sirlin}}]{beg:77}%
  \BibitemOpen
  \bibfield  {author} {\bibinfo {author} {\bibfnamefont {M.~A.~B.}\
  \bibnamefont {B\'eg}}, \bibinfo {author} {\bibfnamefont {R.~V.}\ \bibnamefont
  {Budny}}, \bibinfo {author} {\bibfnamefont {R.}~\bibnamefont {Mohapatra}}, \
  and\ \bibinfo {author} {\bibfnamefont {A.}~\bibnamefont {Sirlin}},\ }\href
  {\doibase 10.1103/PhysRevLett.38.1252} {\bibfield  {journal} {\bibinfo
  {journal} {Phys. Rev. Lett.}\ }\textbf {\bibinfo {volume} {38}},\ \bibinfo
  {pages} {1252} (\bibinfo {year} {1977})}\BibitemShut {NoStop}%
\bibitem [{\citenamefont {{Paul Langacker}}\ and\ \citenamefont {{S. Uma
  Sankar}}(1989)}]{Langacker:89}%
  \BibitemOpen
  \bibfield  {author} {\bibinfo {author} {\bibnamefont {{Paul Langacker}}}\
  and\ \bibinfo {author} {\bibnamefont {{S. Uma Sankar}}},\ }\href {\doibase
  10.1103/PhysRevD.40.1569} {\bibfield  {journal} {\bibinfo  {journal} {Phys.
  Rev. D}\ }\textbf {\bibinfo {volume} {40}},\ \bibinfo {pages} {1569}
  (\bibinfo {year} {1989})}\BibitemShut {NoStop}%
\bibitem [{\citenamefont {Mohapatra}\ and\ \citenamefont
  {Senjanovi\ifmmode~\acute{c}\else \'{c}\fi{}}(1980)}]{mohapatra:80}%
  \BibitemOpen
  \bibfield  {author} {\bibinfo {author} {\bibfnamefont {R.~N.}\ \bibnamefont
  {Mohapatra}}\ and\ \bibinfo {author} {\bibfnamefont {G.}~\bibnamefont
  {Senjanovi\ifmmode~\acute{c}\else \'{c}\fi{}}},\ }\href {\doibase
  10.1103/PhysRevLett.44.912} {\bibfield  {journal} {\bibinfo  {journal} {Phys.
  Rev. Lett.}\ }\textbf {\bibinfo {volume} {44}},\ \bibinfo {pages} {912}
  (\bibinfo {year} {1980})}\BibitemShut {NoStop}%
\bibitem [{\citenamefont {Branco}\ \emph {et~al.}(1983)\citenamefont {Branco},
  \citenamefont {Fr\`ere},\ and\ \citenamefont {G\'erard}}]{Branco}%
  \BibitemOpen
  \bibfield  {author} {\bibinfo {author} {\bibfnamefont {G.}~\bibnamefont
  {Branco}}, \bibinfo {author} {\bibfnamefont {J.-M.}\ \bibnamefont {Fr\`ere}},
  \ and\ \bibinfo {author} {\bibfnamefont {J.-M.}\ \bibnamefont {G\'erard}},\
  }\href {\doibase https://doi.org/10.1016/0550-3213(83)90581-3} {\bibfield
  {journal} {\bibinfo  {journal} {Nuclear Physics B}\ }\textbf {\bibinfo
  {volume} {221}},\ \bibinfo {pages} {317 } (\bibinfo {year}
  {1983})}\BibitemShut {NoStop}%
\bibitem [{\citenamefont {Carnoy}\ \emph {et~al.}(1988)\citenamefont {Carnoy},
  \citenamefont {Deutsch},\ and\ \citenamefont {{Barry R.
  Holstein}}}]{Carnoy:PRD}%
  \BibitemOpen
  \bibfield  {author} {\bibinfo {author} {\bibfnamefont {A.-S.}\ \bibnamefont
  {Carnoy}}, \bibinfo {author} {\bibfnamefont {J.}~\bibnamefont {Deutsch}}, \
  and\ \bibinfo {author} {\bibnamefont {{Barry R. Holstein}}},\ }\href
  {\doibase 10.1103/PhysRevD.38.1636} {\bibfield  {journal} {\bibinfo
  {journal} {Phys. Rev. D}\ }\textbf {\bibinfo {volume} {38}},\ \bibinfo
  {pages} {1636} (\bibinfo {year} {1988})}\BibitemShut {NoStop}%
\bibitem [{\citenamefont {Naviliat-Cuncic}\ \emph {et~al.}(1991)\citenamefont
  {Naviliat-Cuncic}, \citenamefont {Girard}, \citenamefont {Deutsch},\ and\
  \citenamefont {Severijns}}]{oscar:91}%
  \BibitemOpen
  \bibfield  {author} {\bibinfo {author} {\bibfnamefont {O.}~\bibnamefont
  {Naviliat-Cuncic}}, \bibinfo {author} {\bibfnamefont {T.~A.}\ \bibnamefont
  {Girard}}, \bibinfo {author} {\bibfnamefont {J.}~\bibnamefont {Deutsch}}, \
  and\ \bibinfo {author} {\bibfnamefont {N.}~\bibnamefont {Severijns}},\ }\href
  {http://stacks.iop.org/0954-3899/17/i=6/a=013} {\bibfield  {journal}
  {\bibinfo  {journal} {Journal of Physics G: Nuclear and Particle Physics}\
  }\textbf {\bibinfo {volume} {17}},\ \bibinfo {pages} {919} (\bibinfo {year}
  {1991})}\BibitemShut {NoStop}%
\bibitem [{\citenamefont {Fenker}\ \emph {et~al.}(2018)\citenamefont {Fenker},
  \citenamefont {Gorelov}, \citenamefont {Melconian}, \citenamefont {Behr},
  \citenamefont {Anholm}, \citenamefont {Ashery}, \citenamefont {Behling},
  \citenamefont {Cohen}, \citenamefont {Craiciu}, \citenamefont {Gwinner},
  \citenamefont {McNeil}, \citenamefont {Mehlman}, \citenamefont {Olchanski},
  \citenamefont {Shidling}, \citenamefont {Smale},\ and\ \citenamefont
  {Warner}}]{fenker:18}%
  \BibitemOpen
  \bibfield  {author} {\bibinfo {author} {\bibfnamefont {B.}~\bibnamefont
  {Fenker}}, \bibinfo {author} {\bibfnamefont {A.}~\bibnamefont {Gorelov}},
  \bibinfo {author} {\bibfnamefont {D.}~\bibnamefont {Melconian}}, \bibinfo
  {author} {\bibfnamefont {J.~A.}\ \bibnamefont {Behr}}, \bibinfo {author}
  {\bibfnamefont {M.}~\bibnamefont {Anholm}}, \bibinfo {author} {\bibfnamefont
  {D.}~\bibnamefont {Ashery}}, \bibinfo {author} {\bibfnamefont {R.~S.}\
  \bibnamefont {Behling}}, \bibinfo {author} {\bibfnamefont {I.}~\bibnamefont
  {Cohen}}, \bibinfo {author} {\bibfnamefont {I.}~\bibnamefont {Craiciu}},
  \bibinfo {author} {\bibfnamefont {G.}~\bibnamefont {Gwinner}}, \bibinfo
  {author} {\bibfnamefont {J.}~\bibnamefont {McNeil}}, \bibinfo {author}
  {\bibfnamefont {M.}~\bibnamefont {Mehlman}}, \bibinfo {author} {\bibfnamefont
  {K.}~\bibnamefont {Olchanski}}, \bibinfo {author} {\bibfnamefont {P.~D.}\
  \bibnamefont {Shidling}}, \bibinfo {author} {\bibfnamefont {S.}~\bibnamefont
  {Smale}}, \ and\ \bibinfo {author} {\bibfnamefont {C.~L.}\ \bibnamefont
  {Warner}},\ }\href {\doibase 10.1103/PhysRevLett.120.062502} {\bibfield
  {journal} {\bibinfo  {journal} {Phys. Rev. Lett.}\ }\textbf {\bibinfo
  {volume} {120}},\ \bibinfo {pages} {062502} (\bibinfo {year}
  {2018})}\BibitemShut {NoStop}%
\bibitem [{pdg()}]{pdg1}%
  \BibitemOpen
  \href@noop {} {}\bibinfo {howpublished}
  {\url{http://pdg.lbl.gov/}}\BibitemShut {NoStop}%
\bibitem [{\citenamefont {{M. Tanabashi {\it et al.} (Particle Data
  Group)}}(2018)}]{pdg2}%
  \BibitemOpen
  \bibfield  {author} {\bibinfo {author} {\bibnamefont {{M. Tanabashi {\it et
  al.} (Particle Data Group)}}},\ }\href@noop {} {\bibfield  {journal}
  {\bibinfo  {journal} {Phys. Rev. D}\ }\textbf {\bibinfo {volume} {98}},\
  \bibinfo {pages} {030001} (\bibinfo {year} {2018})}\BibitemShut {NoStop}%
\bibitem [{\citenamefont {{Combs}}\ \emph {et~al.}(2016)\citenamefont
  {{Combs}}, \citenamefont {{Calaprice}}, \citenamefont {{Jones}},\ and\
  \citenamefont {{Young}}}]{combs}%
  \BibitemOpen
  \bibfield  {author} {\bibinfo {author} {\bibfnamefont {D.}~\bibnamefont
  {{Combs}}}, \bibinfo {author} {\bibfnamefont {F.}~\bibnamefont
  {{Calaprice}}}, \bibinfo {author} {\bibfnamefont {G.}~\bibnamefont
  {{Jones}}}, \ and\ \bibinfo {author} {\bibfnamefont {A.}~\bibnamefont
  {{Young}}},\ }in\ \href@noop {} {\emph {\bibinfo {booktitle} {APS Division of
  Nuclear Physics Meeting Abstracts}}}\ (\bibinfo {year} {2016})\ p.\ \bibinfo
  {pages} {KG.008}\BibitemShut {NoStop}%
\bibitem [{you()}]{young}%
  \BibitemOpen
  \href@noop {} {}\bibinfo {howpublished} {A. R. Young, private
  communication.}\BibitemShut {Stop}%
\bibitem [{\citenamefont {Tishchenko}\ \emph {et~al.}(2013)\citenamefont
  {Tishchenko} \emph {et~al.}}]{mulan}%
  \BibitemOpen
  \bibfield  {author} {\bibinfo {author} {\bibfnamefont {V.}~\bibnamefont
  {Tishchenko}} \emph {et~al.} (\bibinfo {collaboration} {MuLan
  Collaboration}),\ }\href {\doibase 10.1103/PhysRevD.87.052003} {\bibfield
  {journal} {\bibinfo  {journal} {Phys. Rev. D}\ }\textbf {\bibinfo {volume}
  {87}},\ \bibinfo {pages} {052003} (\bibinfo {year} {2013})}\BibitemShut
  {NoStop}%
\bibitem [{\citenamefont {Marciano}\ and\ \citenamefont
  {Sirlin}(2006)}]{marciano:06}%
  \BibitemOpen
  \bibfield  {author} {\bibinfo {author} {\bibfnamefont {W.~J.}\ \bibnamefont
  {Marciano}}\ and\ \bibinfo {author} {\bibfnamefont {A.}~\bibnamefont
  {Sirlin}},\ }\href {\doibase 10.1103/PhysRevLett.96.032002} {\bibfield
  {journal} {\bibinfo  {journal} {Phys. Rev. Lett.}\ }\textbf {\bibinfo
  {volume} {96}},\ \bibinfo {pages} {032002} (\bibinfo {year}
  {2006})}\BibitemShut {NoStop}%
\bibitem [{\citenamefont {{Chien-Yeah Seng}}\ \emph {et~al.}(2018)\citenamefont
  {{Chien-Yeah Seng}}, \citenamefont {{Mikhail Gorchtein}}, \citenamefont
  {{Hiren H. Patel}},\ and\ \citenamefont {{Michael J.
  Ramsey-Musolf}}}]{Seng:2018}%
  \BibitemOpen
  \bibfield  {author} {\bibinfo {author} {\bibnamefont {{Chien-Yeah Seng}}},
  \bibinfo {author} {\bibnamefont {{Mikhail Gorchtein}}}, \bibinfo {author}
  {\bibnamefont {{Hiren H. Patel}}}, \ and\ \bibinfo {author} {\bibnamefont
  {{Michael J. Ramsey-Musolf}}},\ }\href {\doibase
  10.1103/PhysRevLett.121.241804} {\bibfield  {journal} {\bibinfo  {journal}
  {Phys. Rev. Lett.}\ }\textbf {\bibinfo {volume} {121}},\ \bibinfo {pages}
  {241804} (\bibinfo {year} {2018})}\BibitemShut {NoStop}%
\bibitem [{\citenamefont {Haeberli}\ and\ \citenamefont {{Barry R.
  Holstein}}(1995)}]{haeberli}%
  \BibitemOpen
  \bibfield  {author} {\bibinfo {author} {\bibfnamefont {W.}~\bibnamefont
  {Haeberli}}\ and\ \bibinfo {author} {\bibnamefont {{Barry R. Holstein}}},\
  }in\ \href@noop {} {\emph {\bibinfo {booktitle} {Symmetries and Fundamental
  Interactions in Nuclei}}},\ \bibinfo {editor} {edited by\ \bibinfo {editor}
  {\bibfnamefont {W.}~\bibnamefont {Haxton}}\ and\ \bibinfo {editor}
  {\bibfnamefont {E.}~\bibnamefont {Henley}}}\ (\bibinfo  {publisher} {World
  Scientific},\ \bibinfo {year} {1995})\ p.~\bibinfo {pages} {17}\BibitemShut
  {NoStop}%
\bibitem [{\citenamefont {Elsener}\ \emph {et~al.}(1984)\citenamefont
  {Elsener}, \citenamefont {Gr\"uebler}, \citenamefont {K\"onig}, \citenamefont
  {Schmelzbach}, \citenamefont {Ulbricht}, \citenamefont {Singy}, \citenamefont
  {Forstner}, \citenamefont {Zhang},\ and\ \citenamefont {Vuaridel}}]{elsener}%
  \BibitemOpen
  \bibfield  {author} {\bibinfo {author} {\bibfnamefont {K.}~\bibnamefont
  {Elsener}}, \bibinfo {author} {\bibfnamefont {W.}~\bibnamefont {Gr\"uebler}},
  \bibinfo {author} {\bibfnamefont {V.}~\bibnamefont {K\"onig}}, \bibinfo
  {author} {\bibfnamefont {P.~A.}\ \bibnamefont {Schmelzbach}}, \bibinfo
  {author} {\bibfnamefont {J.}~\bibnamefont {Ulbricht}}, \bibinfo {author}
  {\bibfnamefont {D.}~\bibnamefont {Singy}}, \bibinfo {author} {\bibfnamefont
  {C.}~\bibnamefont {Forstner}}, \bibinfo {author} {\bibfnamefont {W.~Z.}\
  \bibnamefont {Zhang}}, \ and\ \bibinfo {author} {\bibfnamefont
  {B.}~\bibnamefont {Vuaridel}},\ }\href {\doibase 10.1103/PhysRevLett.52.1476}
  {\bibfield  {journal} {\bibinfo  {journal} {Phys. Rev. Lett.}\ }\textbf
  {\bibinfo {volume} {52}},\ \bibinfo {pages} {1476} (\bibinfo {year}
  {1984})}\BibitemShut {NoStop}%
\bibitem [{\citenamefont {{Bertrand Desplanques}}\ \emph
  {et~al.}(1980)\citenamefont {{Bertrand Desplanques}}, \citenamefont {{John F.
  Donoghue}},\ and\ \citenamefont {{Barry R. Holstein}}}]{DDH}%
  \BibitemOpen
  \bibfield  {author} {\bibinfo {author} {\bibnamefont {{Bertrand
  Desplanques}}}, \bibinfo {author} {\bibnamefont {{John F. Donoghue}}}, \ and\
  \bibinfo {author} {\bibnamefont {{Barry R. Holstein}}},\ }\href {\doibase
  https://doi.org/10.1016/0003-4916(80)90217-1} {\bibfield  {journal} {\bibinfo
   {journal} {Annals of Physics}\ }\textbf {\bibinfo {volume} {124}},\ \bibinfo
  {pages} {449 } (\bibinfo {year} {1980})}\BibitemShut {NoStop}%
\bibitem [{\citenamefont {{Joseph Wasem}}(2012)}]{wasem}%
  \BibitemOpen
  \bibfield  {author} {\bibinfo {author} {\bibnamefont {{Joseph Wasem}}},\
  }\href {\doibase 10.1103/PhysRevC.85.022501} {\bibfield  {journal} {\bibinfo
  {journal} {Phys. Rev. C}\ }\textbf {\bibinfo {volume} {85}},\ \bibinfo
  {pages} {022501} (\bibinfo {year} {2012})}\BibitemShut {NoStop}%
\bibitem [{\citenamefont {Phillips}\ \emph {et~al.}(2015)\citenamefont
  {Phillips}, \citenamefont {Samart},\ and\ \citenamefont {Schat}}]{philips}%
  \BibitemOpen
  \bibfield  {author} {\bibinfo {author} {\bibfnamefont {D.~R.}\ \bibnamefont
  {Phillips}}, \bibinfo {author} {\bibfnamefont {D.}~\bibnamefont {Samart}}, \
  and\ \bibinfo {author} {\bibfnamefont {C.}~\bibnamefont {Schat}},\ }\href
  {\doibase 10.1103/PhysRevLett.114.062301} {\bibfield  {journal} {\bibinfo
  {journal} {Phys. Rev. Lett.}\ }\textbf {\bibinfo {volume} {114}},\ \bibinfo
  {pages} {062301} (\bibinfo {year} {2015})}\BibitemShut {NoStop}%
\bibitem [{\citenamefont {Schindler}\ \emph {et~al.}(2016)\citenamefont
  {Schindler}, \citenamefont {Springer},\ and\ \citenamefont
  {Vanasse}}]{schindler}%
  \BibitemOpen
  \bibfield  {author} {\bibinfo {author} {\bibfnamefont {M.~R.}\ \bibnamefont
  {Schindler}}, \bibinfo {author} {\bibfnamefont {R.~P.}\ \bibnamefont
  {Springer}}, \ and\ \bibinfo {author} {\bibfnamefont {J.}~\bibnamefont
  {Vanasse}},\ }\href {\doibase 10.1103/PhysRevC.93.025502} {\bibfield
  {journal} {\bibinfo  {journal} {Phys. Rev. C}\ }\textbf {\bibinfo {volume}
  {93}},\ \bibinfo {pages} {025502} (\bibinfo {year} {2016})}\BibitemShut
  {NoStop}%
\bibitem [{\citenamefont {{Wick C. Haxton}}\ and\ \citenamefont {{Barry R.
  Holstein}}(2013)}]{haxton:13}%
  \BibitemOpen
  \bibfield  {author} {\bibinfo {author} {\bibnamefont {{Wick C. Haxton}}}\
  and\ \bibinfo {author} {\bibnamefont {{Barry R. Holstein}}},\ }\href
  {\doibase https://doi.org/10.1016/j.ppnp.2013.03.009} {\bibfield  {journal}
  {\bibinfo  {journal} {Progress in Particle and Nuclear Physics}\ }\textbf
  {\bibinfo {volume} {71}},\ \bibinfo {pages} {185 } (\bibinfo {year}
  {2013})}\BibitemShut {NoStop}%
\bibitem [{\citenamefont {{Susan Gardner}}\ \emph {et~al.}(2017)\citenamefont
  {{Susan Gardner}}, \citenamefont {Haxton},\ and\ \citenamefont {{Barry R.
  Holstein}}}]{gardner}%
  \BibitemOpen
  \bibfield  {author} {\bibinfo {author} {\bibnamefont {{Susan Gardner}}},
  \bibinfo {author} {\bibfnamefont {W.~C.}\ \bibnamefont {Haxton}}, \ and\
  \bibinfo {author} {\bibnamefont {{Barry R. Holstein}}},\ }\href {\doibase
  10.1146/annurev-nucl-041917-033231} {\bibfield  {journal} {\bibinfo
  {journal} {Ann. Rev. Nucl. Part. Sci}\ }\textbf {\bibinfo {volume} {67}},\
  \bibinfo {pages} {69} (\bibinfo {year} {2017})}\BibitemShut {NoStop}%
\end{thebibliography}%

\end{document}